\documentclass[pra,twocolumn,amsmath,amssymb,aps]{revtex4-2}

\usepackage{xcolor}
\usepackage{soul}

\usepackage{graphicx}% Include figure files
\usepackage{dcolumn}% Align table columns on decimal point
\usepackage{bm}% bold math
\usepackage{changepage}%manipulate margin
\usepackage{hyperref}
\usepackage{natbib}
\usepackage{comment}

\begin{document}

\title{Bifurcation structure of  traveling  pulses  in  Type-I  excitable  media} 

\author{Pablo Moreno-Spiegelberg$^1$}
\author{Andreu Arinyo-i-Prats$^{2}$}
\author{Daniel Ruiz-Reyn\'es$^3$}
\author{Manuel A. Matias$^1$}
\author{Dami\`a Gomila$^1$}
\email{damia@ifisc.uib-csic.es}
\affiliation{$^1$IFISC (CSIC-UIB), Instituto de F\'{\i}sica Interdisciplinar y Sistemas Complejos,  E-07122 Palma de Mallorca,  Spain \\
$^2$ Institute of Computer Science, Czech Academy of Sciences, 182 07 Prague 8, Czech Republic\\
$^3$ Laboratory of Dynamics in Biological Systems, Department of Cellular and Molecular Medicine, University of Leuven (KU Leuven), B-3000 Leuven, Belgium.
}
\date{\today}

\begin{abstract}
We study the scenario in which traveling pulses emerge in a prototypical Type-I $1-$dimensional excitable medium, that exhibits two different routes to excitable behavior, mediated by a homoclinic (saddle-loop) and a Saddle-Node on the Invariant Cycle (SNIC) bifurcations.
We characterize the region in parameter space in which traveling pulses are stable together with the different bifurcations behind either their destruction or loss of stability. In particular, some of the bifurcations delimiting the stability region have been connected, using singular limits, with the two different scenarios that mediated Type-I local excitability. Finally, the existence of traveling pulses has been linked to a drift pitchfork instability of localized steady structures.
\end{abstract}
\maketitle

\section{Introduction}

Excitable dynamical systems have a linearly stable rest state (i.e. a fixed point) that, under nonlinear perturbations above a certain amplitude, experience a long excursion in phase space.
Excitable media are dynamical systems extended in space that are locally excitable, and so perturbations of the rest state may propagate in the system. This is because a local perturbation that excites an excursion in phase space can trigger an excitable excursion of the nearest neighbor points, to finally return back to the rest state.  
In 1-D, the result of this process is the creation of a traveling pulse (TP). After a pulses passes through, the system returns (locally) to the rest state, becoming susceptible to be excited again and, therefore, TPs can pass many times through the same region, leading to a huge variety of spatio-temporal structures, such as solitary pulses, wave trains, and, in $2-$D, target patterns, or rotating spirals \cite{Meron92,Mikhailovbook,Kapral95,Alonso2016}. 
Two ingredients are essential to observe these dynamic regimes. On one hand, temporal excitability of the local dynamics. On the other hand, a spatial coupling between the elements of the system to allow a perturbation to propagate in space.

Temporal excitability is usually associated with the sudden destruction of a large amplitude limit cycle.
Remnant traces of this cycle in phase space constitute the excitable excursion.
The bifurcation through which the limit cycle is destroyed leads to differences in the excitable trajectory and, therefore, unique properties of the excitable system.
Temporal excitability can be classified into two types, I and II, depending on which kind of bifurcation mediates the transition to the limit cycle \cite{ErmenRinzel,Izhikevich2000,IzhikevichDSN}.

Type-I excitability is generated at a Saddle-Node on Invariant Cycle ($SNIC$) or Homoclinic (also known as saddle-loop) bifurcations, involving a saddle point aside from the stable rest point \cite{IzhikevichDSN}.
It has been profusely reported in Neuroscience \cite{IzhikevichDSN} and also in other fields \cite{Plaza1997, Dubbeldam1999,Gomila2005}.
More generally, Type-I excitable behaviors can be found in the case that the excitable system is defined in a higher-dimensional phase space, as is the case of 3-variable excitable systems exhibiting a homoclinic to a saddle-focus, instead of a saddle, i.e. a Shilnikov scenario, that leads to multi-pulse excitability \cite{Wieczorek02,Yochelis2008a,Yochelis2015}.
In turn, Type-II excitability is well explained in textbooks with a Neuroscience orientation \cite{IzhikevichDSN,GerstnerBook} and is the type of excitability found in the well known FitzHugh-Nagumo model.
Type-II is mediated by a Hopf bifurcation that creates very large stable cycle in a very narrow parameter space.
Typically, this corresponds to a supercritical Hopf followed by a canard \cite{Benoit1981}, i.e. a sudden growth of the cycle happening in fast-slow systems, or to
a subcritical Hopf with a fold of cycles.
In turn, each bifurcation confers to Type-I and Type-II excitabilities unique and distinct features.

In general, in Type-I excitability: i) the distance between the stable rest point and the saddle defines a threshold in phase space, given by the stable manifold of the saddle;  ii) the duration of an excitable excursion depends on how close the initial condition is to the threshold, diverging for initial conditions on it. 

In Type-II excitability instead: i) there is not a well defined threshold in phase space, but a narrow region of initial condition produces a pseudothreshold. The thickness of this region depends on the ratio between the time scales of the system. Hence, the excitability is only well defined when the difference in time scales is large enough. ii) The differences in temporal scales provide trajectories with a well defined duration determined by the slow dynamics, independently of the initial condition.

We will coin Type-I (II) excitable media to the space extended system obtained by adding diffusion to a Type-I (II) excitable system. Since Type-I and Type-II excitability are intrinsically different, it is expected to find some differences between Type-I and II excitable media. The traveling pulses found in both kinds of media share some general features. In both cases, TPs are isolated, having speed and shape not determined by the initial conditions, and annihilate when collide with other pulses. Nevertheless, the mechanisms behind the creation and stabilization of the TP (and the behavior of the pulse close to these points) depend on the associated temporal excitability. Also the shape of TP may quite differ depending on the type of excitability. For instance, Type-II excitable pulses in systems with a big time scale separation have a distinctive square-like form which is not observed in the Type-I case.  

Until quite recently, most studies of excitable media have been carried out in the case that the excitable medium is locally of Type-II, using models that have the corresponding bifurcations in the local dynamics \cite{Meron92,Mikhailovbook}.
These include examples in neuroscience \cite{follmann2015dynamics}, depression waves \cite{dahlem2004reaction}, cardiac tissue \cite{Alonso2016},  reaction-diffusion systems \cite{lee1994experimental}, and nonlinear optics \cite{Marino05}. The origin of the TPs in these systems can be tracked to the singular limit when the difference between time scales diverges \cite{Meron92,Mikhailovbook}. This mechanism is strongly related with the bifurcations associated with the Type-II excitability of the local part.

Nevertheless, there is a number of recent studies in the context of vegetation dynamics \cite{ ruiz2017fairy, Oto2019, zhao2021,Ruiz-Reynes2022} that report excitable behavior where the local dynamics exhibits features indicating that they are of Type-I. Similar behaviors have also been found in other biological systems \cite{follmann2015dynamics,Romeo2003}. In some of these systems the origin of the TP is associated to a T point bifurcation \cite{Romeo2003,Champneys2007}. However, the relation between the temporal (local) bifurcations of the system, and its spatial structure counterpart (in the case of TPs) has not been properly analyzed in the literature. Connecting a spatially extended system to its local dynamics is a powerful approach. This concept has been successfully applied to explain the emergence of chaos, stationary pulses and kink patterns in conserved reaction diffusion systems \cite{Halatek2018,Brauns2020}.

In \cite{arinyoiprats2021traveling} we discussed the existence and properties of $1$-D TPs in the simplest general model for Type-I excitable media we are aware of, that exhibits both $SNIC$ and homoclinic bifurcations, and showed how the shape of TPs was affected by the proximity in the parameter space to the bifurcations leading to Type-I excitability. In this work, we extend this analysis by describing the full bifurcation diagram and dynamical regimes of TPs in Type-I excitable media. First, we fully characterize the different bifurcations delimiting the stable region of excitable traveling pulses. 
We explicitly show the connection between some of the bifurcations of the  temporal system and the spatio-temporal dynamics, as was suggested in \cite{arinyoiprats2021traveling}. 
Furthermore, we investigate and characterize bifurcations that are intrinsically spatiotemporal, i.e. with no purely temporal counterpart. 
Finally, the existence of TPs has been tracked outside the excitable region, until their creation point, connecting excitable traveling pulses with a drift pitchfork bifurcation of localized steady solutions of bistable systems. 

The paper is structured as follows: In Section \ref{model} we describe the model we analyze. In Section \ref{TPSR} we characterize the excitable pulses, we introduce its stability region, and show the scenarios found when abandoning this region through the different bifurcations. In Section \ref{MSDS_bif} we analyze in detail the bifurcations that delimited the stability region and connect some of them with spaceless counterparts. In Section \ref{DP} we find that the traveling pulses emerge, outside the stability region, from localized steady structures through a Drift Pitchfork bifurcation. Finally, in Section \ref{Conclusions} we give some concluding remarks.

\section{Model} \label{model}

We consider a general reaction-diffusion model given by the following system of partial differential equations (PDEs):
\begin{align}
\label{Spsist}
\partial_t u &=v + \partial_{xx} u  \\
\partial_t v &= -u^{3}+\mu _{2} u+\mu _{1} +v(\nu + bu -  u^{2}) + \partial_{xx} v\ . \nonumber
\end{align}
The local dynamics of the system is described by the bounded normal form of a codimension-3 degenerate Takens-Bogdanov bifurcation with triple equilibrium \cite{dumortier2006bifurcations}. This is the simplest continuous model which gives a complete description of Type-I excitability, in the sense that the excitable region is accessible either through a homoclinic or a $SNIC$ bifurcation. We have added $1$-D diagonal diffusion to study spatial propagation. The diffusion coefficients have been chosen to be equal in both fields, and without loss of generality, they have been fixed to one. This choice does not introduce any special symmetry, nor generates spatial instabilities of the homogeneous solutions.  We have also checked that slightly different diffusion constants in each field does not substantially modify the results of this work. We fix the parameters $\nu = 1 $, $b = 2.4$ and consider $\mu_{1}$ and $\mu_{2}$ as control parameters \cite{arinyoiprats2021traveling}.

Excitable pulses in PDEs can be studied using two different approaches.
The first approach is the local dynamical system, which describes the evolution of the system without diffusion or, alternatively, the evolution of homogeneous solutions of the system.
Excitable pulses somehow transcribe local dynamics in space, resembling in their profile the excitable trajectory in time \cite{arinyoiprats2021traveling}.
Furthermore, the bifurcation behind the creation of excitable trajectories in the local dynamical system is connected with the creation of excitable pulses in the PDEs, as will be expanded in Section~\ref{MSDS_bif}.
The second approach is the moving spatial dynamical system, which describes the spatial profile of structures that propagates with a fixed velocity $c$ without changing shape.
From this approach, TPs are interpreted as trajectories in a system of ordinary differential equations.
Therefore, the creation of TPs can be studied through the bifurcations that generate homoclinic trajectories.
In the limit $c \rightarrow \infty$, the local dynamical equations are recovered from the moving spatial dynamical system, as shown in Section~\ref{MSDS}, connecting both approaches.

\subsection{Local dynamics}
The dynamics of the local system is described by the following set of ordinary differential equations:
\begin{align}
\label{Localsist}
\frac{du}{dt} &=v  \\
\frac{dv}{dt} &= -u^{3}+\mu _{2} u+\mu _{1} +v(\nu +bu-  u^{2}) . \nonumber
\end{align}

\begin{figure*}
\centering
\includegraphics[width=0.70\textwidth]{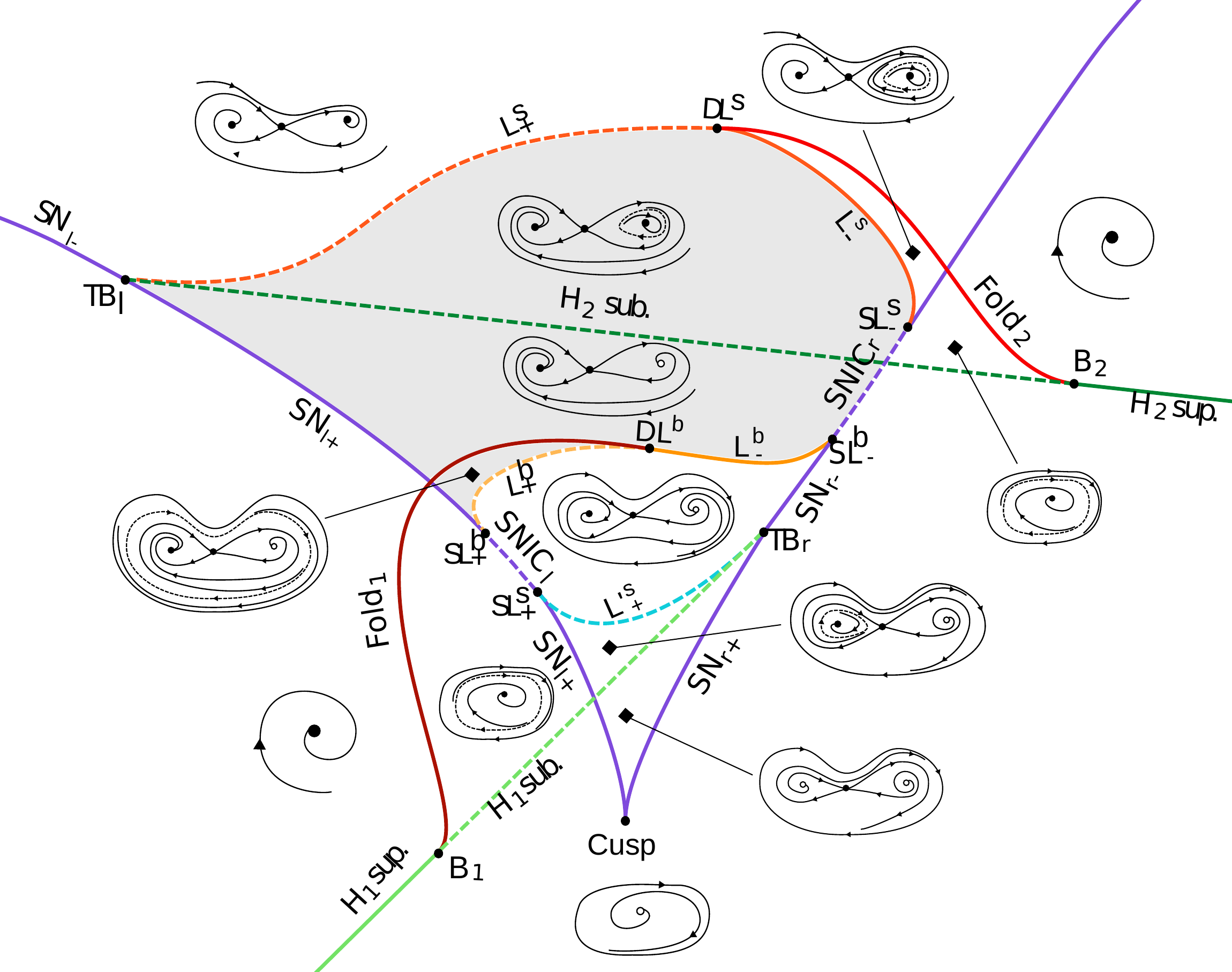}
\caption{\label{fig:diagram} Schematic phase diagram for the $\mu_1$, $\mu_2$ plane for fixed $\nu = 1$, $b = 2.4$ parameters of equation \ref{Localsist}. The lines represent the bifurcations that separate the parameter space in qualitative different behaviors of the system. The dots mark the codimension-2 points. The diagrams show a schematic representation of the phase portrait in the different regions. Gray area marks the excitability region.}
\end{figure*}

\begin{figure*}
\centering
\includegraphics[width=0.70\textwidth]{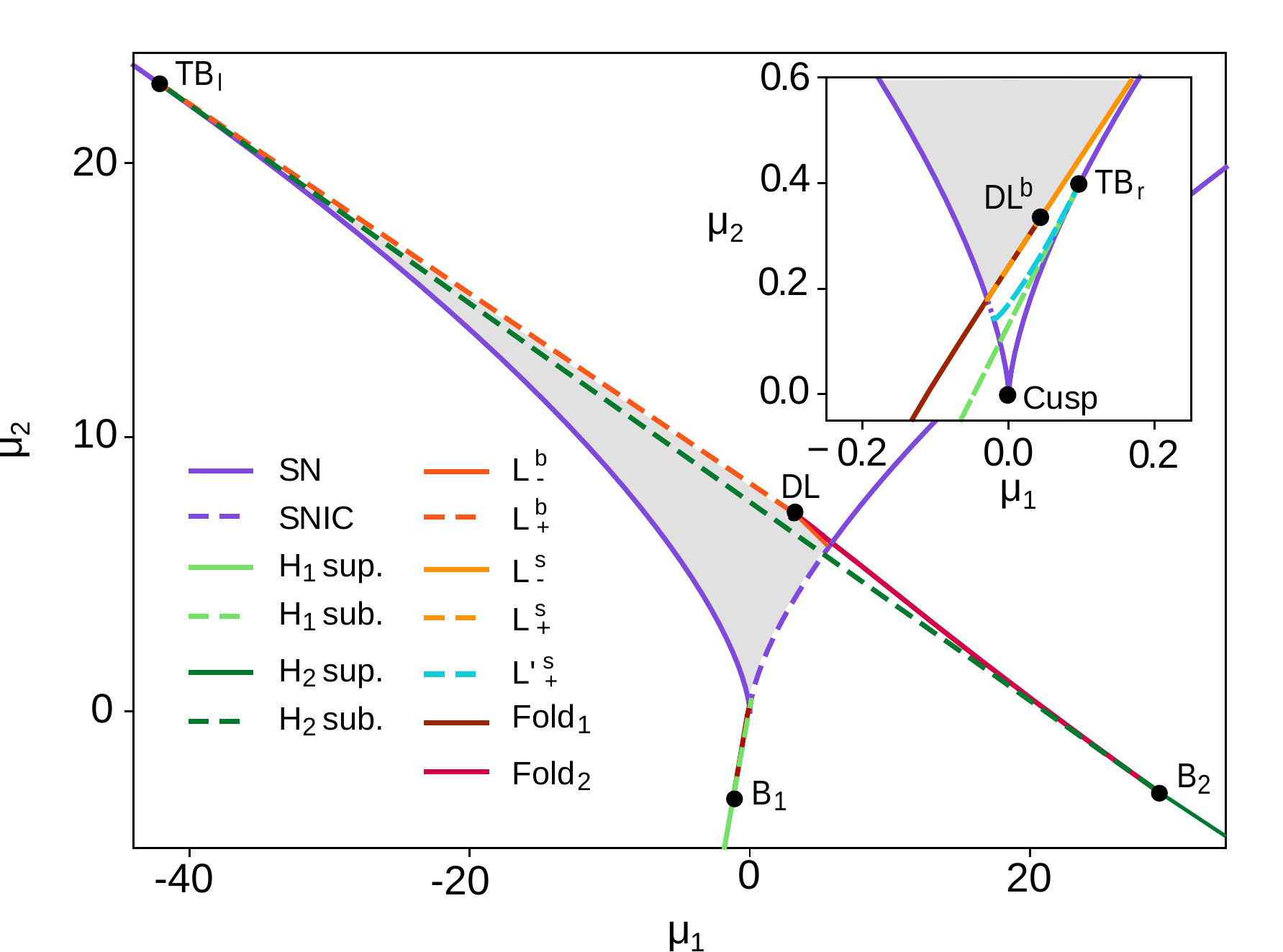}
\caption{\label{fig:saddle} Actual phase diagram of the temporal system (\ref{Localsist}) for $\nu=1$ and $b=2.4$.}
\end{figure*}

A schematic phase diagram showing the different bifurcations and dynamical regions is shown in Fig.~\ref{fig:diagram} for $\nu = 1$ and $b=2.4$. The actual phase diagram is shown in Fig.~\ref{fig:saddle}.

The fixed points of this temporal system are $v^{\star}=0$ and $u^{\star}$ being determined by the solutions to the cubic equation
\begin{align}
    \label{Fixu}
    -u^{*3}+\mu_2u^*+\mu_1 = 0,
\end{align}
 which represents the unfolding of a Cusp bifurcation point, located at $\mu_1=\mu_2=0$ (see Fig. \ref{fig:diagram}). Two saddle-node bifurcation lines start from this point and are given by:
 \begin{align}
\label{eq:saddle}
SN_l:\qquad &\mu_{1}=- 2\sqrt{\frac{\mu_{2}^{3}}{27}} \qquad \mu_2 >0\\
SN_r:\qquad &\mu_{1}=+ 2\sqrt{\frac{\mu_{2}^{3}}{27}} \qquad \mu_2 >0.\nonumber
\end{align}
In Fig.~\ref{fig:diagram} we attach a plus or a minus to the $SN$ labels depending on whether they create two unstable fixed points or a saddle and a stable fixed point.
These $SN$ bifurcations separate the parameter space in two regions. In the inner region, where the cubic has three real roots, the system has three fixed points: $\left\lbrace P_i (u^*_i,v^*=0)~:~i=1,2,3 \right\rbrace$ where $u^*_i$ are the roots of Eq.~(\ref{Fixu}) and $u^*_1<u^*_2<u^*_3$. In the outer region the system has a single fixed point $P_0(u^*_0,v^*=0)$, as $u^*_0$ is the only root of Eq.~(\ref{Fixu}).
  
A linear stability analysis provides the eigenvalues of the fixed points of (\ref{Localsist}), given by:
\begin{align}
    \lambda_{\pm} \left( P_i\right) = \lambda_{\pm} \left( u^*_i\right) = \frac{\tau\left( u^*_i\right) \pm \sqrt{\tau^2\left( u^*_i\right)-4\Delta\left( u^*_i\right)}}{2},
    \label{eq:temp_eig}
\end{align}
where $\tau\left( u^*_i\right)$ and $\Delta\left( u^*_i\right)$ are the trace and determinant of the Jacobian at $P_i$ given by:
\begin{align}
    \tau\left( u^*_i\right) = \nu + b u^*_i -u^{*2}_i \\
    \Delta\left( u^*_i\right) = 3 u^{*2}_i - \mu_2.
\end{align}
Attending to the eigenvalues, the point $P_2$ is always a saddle while $P_0$, $P_1$ and $P_3$ can be, depending on the parameters, either foci or nodes.

On the $SN_l$ ($SN_r$) bifurcation line the fixed points $P_2$ and $P_3$ ($P_2$ and $P_1$) collide in a  Saddle-Node point and are annihilated. The point $P_1$ ($P_3$) is not affected by this bifurcation, and it is renamed to $P_0$ in the outside region.

For the values of $\nu$ and $b$ used in this paper, the solutions $P_{0,1,3}$ can change their stability through Andronov-Hopf bifurcations. Two different Hopf bifurcation lines have been found. One affecting $P_1$, label as $H_1$; and another one which involves $P_3$, label as $H_2$. Outside the bistable region, both bifurcations lines affect the stability of $P_0$. 
These bifurcations are located at:
\begin{align}
     \mu_1= u_{H1,2}^3 - \mu_2 u_{H1,2} && \mu_2<3u^2_{H1,2} 
 \label{eq:hopfmu}
\end{align}
where $u_{H1,2}$ is the value of the steady homogeneous solution at the bifurcation:
\begin{align}
 u_{H1,2}=\dfrac{b\pm \sqrt{b^{2}+4\nu}}{2}
 \label{eq:hopfu}
\end{align}
being $u_{H1}$ the one with a minus sign and $u_{H2}$ the one with a plus sign.

These Hopf bifurcations are sub-critical for $\mu_2>6bu_{H1,2}-9u^2_{H1,2}$ and supercritical for $\mu_2<6bu_{H1,2}-9u^2_{H1,2}$.
The point where the Hopf bifurcation change from super to subcritical, $(\mu_1,\mu_2)=(10u^3_{H1,2}-6bu^2_{H1,2},6bu_{H1,2}-9u^2_{H1,2})$, is known as Bautin point ($B$), or also degenerate Hopf.
From each of the Bautin points emerges, tangentially to the Hopf, a Fold of cycles, also known as saddle node of periodic orbits.
Although the Fold of cycles curve can not be analytically found, we have obtained a numerical approximation of several points of the Fold of cycles associated with $H_1$ ($Fold_1$) and the fold associated with $H_2$ ($Fold_2$) as it can be seen in Fig \ref{fig:saddle}.

%___________________figure__________________%
% \begin{figure}
% \centering
% \includegraphics[width=0.49\textwidth]{Fig1_parameter_draw.pdf}
%________________End figure_________________%

%___________________figure__________________%
% \begin{figure}
% \centering
% \includegraphics[width=0.49\textwidth]{Fig2_local_bifurcation_diagram.pdf}
%________________End figure_________________%

The $Fold_1$ line ends in a "big" degenerate loop point ($DL^b$) where the degenerate cycle created at the Fold becomes a Homoclinic connection of the $P_2$ fixed point surrounding $P_1$ and $P_3$. The term big denotes that the created cycle surrounds the three fixed points. From this $DL^b$ point starts, tangentially to $Fold_1$, two big homoclinic bifurcations ($L^{b}_{\pm}$). At the $L^{b}_{+}$ ($L^{b}_{-}$) an unstable (stable) limit cycle is created from a Homoclinic connection of the $P_2$ fixed point, embracing the three fixed points. The $L^{b}_{+}$ ($L^{b}_{-}$) meets tangentially the $SN_l$ ($SN_r$) curve in a Saddle-Node Separatrix-Loop codimension-2 point that we call $SL^{b}_+$ ($SL^{b}_-$).

The $Fold_2$ line ends in a "small" degenerate loop point ($DL^s$) where the degenerate cycle created at the Fold becomes a Homoclinic connection of $P_2$ surrounding $P_3$. The term small denotes that the created cycle surrounds only one fixed point. From this point two small  homoclinic bifurcations ($L^{s}_{\pm}$) start tangentially to the $Fold_2$ line. At the $L^{s}_{+}$ ($L^{s}_{-}$) line an unstable (stable) limit cycle is created from an Homoclinic connection of $P_2$, surrounding just $P_3$. The $L^{s}_-$ line ends in a Saddle-Node Separatrix-Loop codimension 2 point ($SL^{s}_+$) where it meets tangentially the $SN_r$. The $L^{s}_{+}$ curve ends in a Takens-Bogdanov codimension 2 point ($TB_l$) tangentially to the $H_2$ and the $SN_l$ curves.

The $H_1$ ($H_2$) line meets tangentially with the $SN_r$ ($SN_l$) curve in a Takens-Bogdanov codimension 2 point, $TB_r$ ($TB_l$), at:
\begin{align}
     \mu_1= -2u_{H1,2}^3 && \mu_2=3u^2_{H1,2} 
 \label{eq:TB}
\end{align}

From each TB point, a small homoclinic bifurcation starts tangentially to the Hopf and $SN$ lines. At these curves, the cycle created in the Andronov-Hopf bifurcation collapse with the saddle fixed point and is destroyed. On one hand, the homoclinic arising from $TB_r$ ($L^{'s}_-$) ends tangentially to $SN_l$ in Saddle-Node Separatrix-Loop codimension 2 point ($SL^{s}_+$). On the other hand, the homoclinic arising from $TB_l$  correspond with the $L^{s}_+$ curve mentioned before.

Between the points $SL^{b}_+$ and $SL^{s}_+$ ($SL^{b}_-$ and $SL^{s}_-$) the $SN_r$ ($SN_l$) bifurcation line creates a stable (unstable) limit cycle, when crossed in the 1-fixed-point direction, in an infinite period bifurcation known as saddle node on the invariant cycle $SNIC_r$ ($SNIC_l$).

In the region delimited by the $SN_{l+}$, $L^{b}_{\pm}$, $SNIC_r$, and  $L^{s}_{\pm}$, shaded in gray in Figures~\ref{fig:diagram} and \ref{fig:saddle}, the system displays Type-I excitable behavior. 
While the system is in the stable fixed point $P_1$, a perturbation that moves the system away from the fixed point and below the stable manifold of the $P_2$ decays exponentially. However, a perturbation that sets the system above this separatrix, grows, making the system explore the remnants of the cycle on a long excursion in phase space, an excitable trajectory, to return afterwards to the rests state $P_1$.

\subsection{Spatial dynamics in a moving reference frame} \label{MSDS}
To interpret some of the bifurcations involved in the creations of the pulses, it is useful to use a different theoretical framework to complement the PDEs  (\ref{Spsist}). 
As pulses are steady structures that propagate with constant speed $c$, we rewrite these solutions in a co-moving reference frame as a function of a single space-time variable $\xi =x-ct$:
\begin{align}
    (u_p (x,t),v_p(x,t)) = (u_p (\xi=x-ct),v_p(\xi=x-ct)).
\end{align}
Rewriting Eqs.~(\ref{Spsist}) for stationary solutions in the moving reference frame we obtain a moving spatial dynamical system (MSDS)\footnote{The spatial dynamical systems in the moving reference frame is sometimes also referred to as traveling wave ordinary differential equations (TWODEs) \cite{Or-Guil2001}}:
\begin{eqnarray}
\label{Spsistrav}
du/d\xi &=&y\qquad ;\qquad dv/d\xi =z \\
dy/d\xi&=&-v - c\, y  \nonumber \\
dz/d\xi&=&u^{3}-\mu _{2} u-\mu _{1} -v (\nu +bu- u^{2})-c\, z  \nonumber\ .
\end{eqnarray}
Non-trivial bounded trajectories of this system define the spatial shape of structures that propagate without changing shape with velocity $c$ in the spatially extended system (\ref{Spsist}) (e.g. limit cycles of (\ref{Spsistrav}) describe traveling wave solutions, homoclinic connections describe traveling pulses, and heteroclinic connections describe propagating fronts between homogeneous solutions).

We would like to point out some considerations about the MSDS.
First, the MSDS gives straightforward information about the existence and shape of steady solutions of the PDEs in a moving reference frame. It does not provide information about the existence of other kinds of dynamical attractors, such as breathers or turbulent regimes. 

Second, the MSDS has one more parameter than the PDEs, the velocity $c$. The PDEs can show, for a fixed parameter configuration, all the structures of the MSDS for any value of $c$. Therefore, the codimension of regions of existence of any structure is greater by a unit in the MSDS.
For example, structures existing only on codimension-1 bifurcations of the MSDS, as traveling pulses, will be codimension-0 in the PDEs.

As a consequence of the spatial reversibility (and time translation) of (\ref{Spsist}), the MSDS remains invariant under the involution:
\begin{align}
    R:(\xi,u,v,y,z,c)\rightarrow(-\xi,u,v,-y,-z,-c)
    \label{involution}
\end{align}
and, therefore, space reversed traveling solution will propagate with opposite velocity. 

Fixed points of the system (\ref{Spsistrav}) describe the homogeneous solutions of the PDEs,  i.e. the fixed points of the local dynamical system (\ref{Localsist}). Therefore, for simplicity, we will use the same notation for the fixed points of the local dynamical system and the MSDS. Linear stability analysis of these points, as well as local bifurcations of the MSDS system are discussed in Appendix~\ref{local_MSDS}. It is important to notice that the MSDS (\ref{Spsistrav}) is not an excitable system in itself. This is because hypothetical excitations to its fixed points will, in general, not return to the same local stationary state, as should happen in a true excitable system. So, true features of excitability are observed in the local dynamics, not in the MSDS. Excitability manifests itself in the MSDS in the form of the homoclinic trajectories, which inherit properties of the temporal excitable excursions. These homoclinic connections are the excitable traveling pulses of the PDEs.

There are two limit cases of special interest. 
The first case is the limit $c\rightarrow \infty$, which describes the temporal evolution of homogeneous solutions of the PDEs. 
Approaching this limit the evolution of bounded trajectories asymptotically slow down, while they approach the plane $(u,v,y,z: y = z = 0)$.
Nevertheless, the time evolution of each point of the associated PDEs solution, ($\mathbf{u_t} = c \mathbf{u}_{\boldsymbol{\xi}}$) converges to the local evolution of that point given by (\ref{Spsist}).

The second case is $c=0$, which defines the steady solutions of the PDEs.
In this particular case there is a set of points invariant under $R$, given by the plane $\Pi = (u,v,y,z: y = z = 0)$. 
Trajectories that cross $\Pi$ are reversible and, therefore, achiral \cite{WOODS1999,Coullet2000}.

\section{traveling pulse stability region}
\label{TPSR}

\begin{figure}
 \centering
\includegraphics[width=0.49\textwidth]{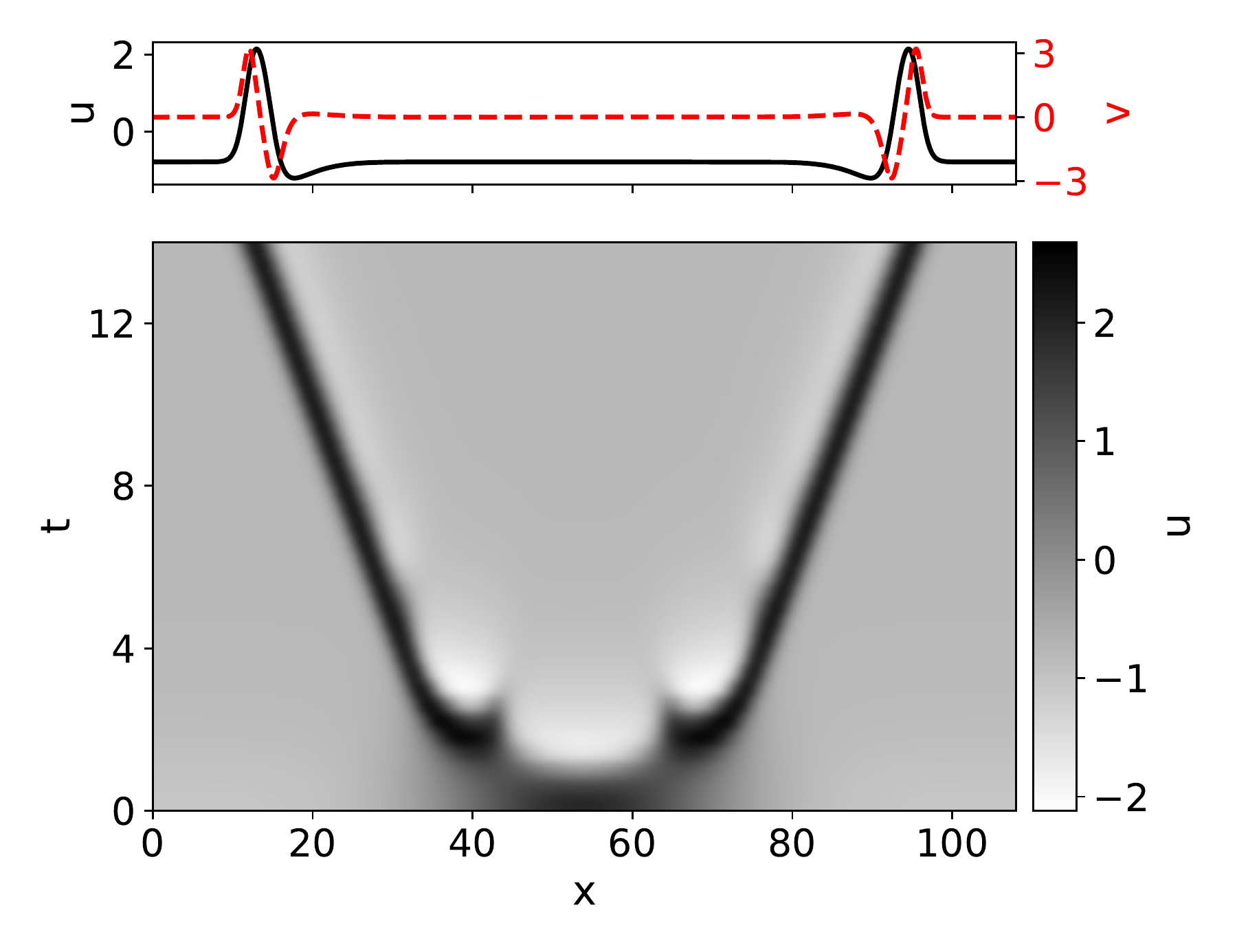}
\caption{\label{fig:gauss_creation} Creation of a pair of traveling pulses from a Gaussian localized initial condition on top of the $P_1$ homogeneous solution. The main figure shows the spatio-temporal evolution of the $u$ field for $\mu_1 = 0.3$ and $\mu_2 = 1.0 $. The top panel shows the transverse profile of the $u$ and $v$ fields for $t=14$.}
\end{figure}

A strong enough localized perturbation of the $P_1$ homogeneous solutions generates, for the appropriate parameter values inside the excitable region, a pair of traveling pulses that propagate in opposite directions in the media (Fig.~\ref{fig:gauss_creation}).
These pulses propagate without changing shape. Two such pulses also cancel each other when colliding. They are, therefore, excitable traveling pulses.
In this paper we will refer to these traveling pulses with the $P_1$ solution as background state as TP.
%___________________figure__________________%
%________________End figure_________________%

In the MSDS, a TP corresponds to a tangential homoclinic connection to the $P_1$ fixed point.
This connection is created at a homoclinic bifurcation, in which limit cycles of the MSDS (waves trains of the PDEs) are destroyed. This bifurcation is codimension-1 in the MSDS.
As a consequence the TPs are isolated and their shape and velocity are independent of the initial conditions in the PDE, thus shape and velocity depend exclusively on the choice of parameters.

\begin{figure*}
 \centering
\includegraphics[width=\textwidth]{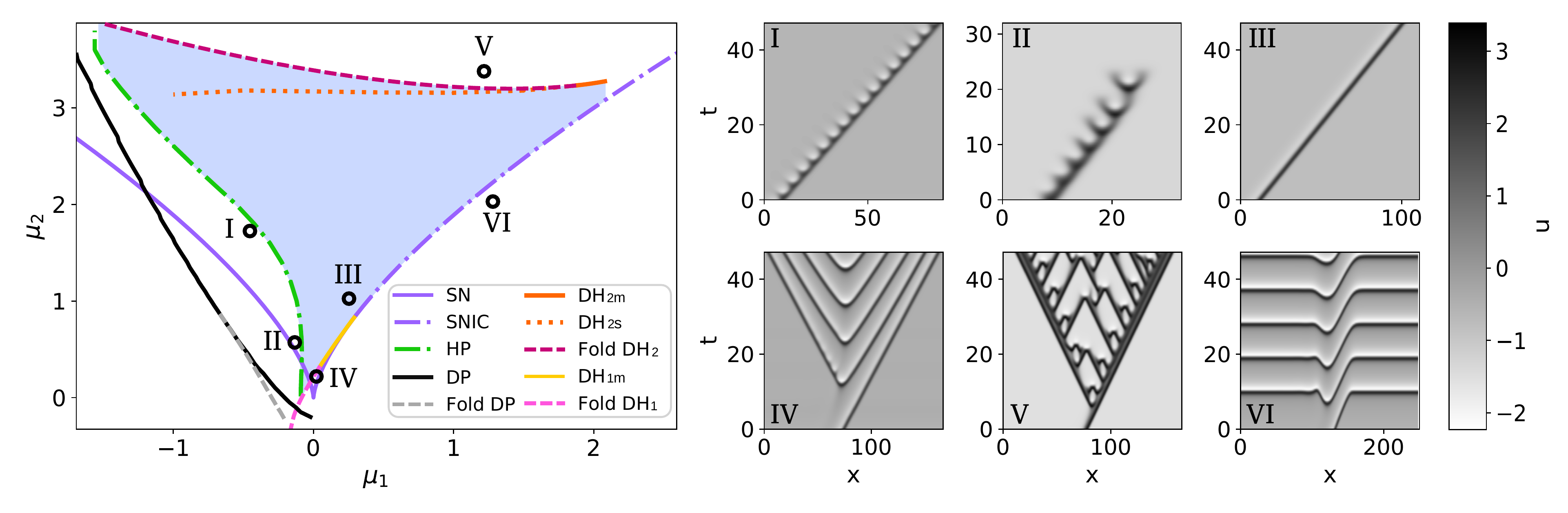}
\caption{\label{fig:stability} Stability region of TP (shaded in blue). The stability region of the pulse is bounded by two monotonic $DH$ point bifurcation, $DH_{1m}$ and $DH_{2m}$, (light and dark orange solid lines), two folds associated with the $DH_{1}$ and $DH_{2}$ bifurcations (pink and magenta dashed lines), Hopf bifurcation of the pulse ($HP$) (green dashed dotted line) and the $SNIC$ (purple dashed dotted line). Also have drew the Drift Pitchfork ($DP$) bifurcation line where the pulses are generated (black line), the fold associate with the $DP$ (gray dashed line), the $DH_{2}$ with collapse snaking bifurcation ($DH_{2s}$, dark orange dotted line)  and the $SN$ bifurcation of the homogeneous solutions as a referential frame (purple lines). Insets I-VI show the dynamics observed for the different parameter values indicated in the main figure. I. Stable oscillatory traveling pulse found once crossed the Hopf bifurcation (\S \ref{subsec:hopfTP}) of TP for $\mu_1 = -0.112$, $\mu_2 = 0.206$. II. Propagation failure found once crossing the Hopf (\S \ref{subsec:hopfTP}) of pulses for $\mu_1 = -0.36$ and $\mu_2 = 1.46 $. III. Stable TP found in the shaded in blue region, for  $\mu_1 = 0.3$, $\mu_2 = 1$. IV. Time evolution of a pulse replication for $\mu_1 = 0$ and $\mu_2 = 0.21$, beyond the Heteroclinic bifurcation I (\S \ref{subsec:hetITP}). V. Time evolution of the system in the turbulent regime found beyond the Heteroclinic bifurcation II (\S \ref{subsec:hetIITP}) for $\mu_1 = 1.29$ to $\mu_2 = 3.21$. VI. Time evolution of a pulse on top of the homogeneous oscillatory state for $\mu_1 = 1.9$ and $\mu_2 = 2.5$ found  beyond the $SNIC$ bifurcation (\S \ref{subsec:snicTP}).}
\end{figure*}

The TP solution is a stable attractor in a part of the temporal excitable region.
This stable part is shown in blue in Fig.~\ref{fig:stability}, and it is limited by four different bifurcations: a Hopf bifurcation of TPs ($HP$, dot-dashed green line) on the left, a $SNIC$ bifurcation (dot-dashed purple line), a double heteroclinic bifurcation ($DH_{1m}$, solid light-orange line), and a fold of TPs (Fold $DH_1$, dashed pink line) on the right, and by a double heteroclinic bifurcation ($DH_{2m}$, solid dark orange line) and a fold (Fold $DH_{2}$, dashed magenta line) on the top. 
In the following sections we discuss each one of these bifurcation lines and the regimes arising when each threshold is crossed. The $SNIC$ and double heteroclinic bifurcation ($DH_{1m}$) were already studied in \citep{arinyoiprats2021traveling}.

%___________________figure__________________%
%________________End figure_________________%

%-----------------Subsec-------------%
\subsection{Hopf bifurcation of traveling pulses}
\label{subsec:hopfTP}
On the left, the stable region is limited by a Hopf bifurcation of the TP (green dot-dashed line in Fig.~\ref{fig:stability}). A linear stability analysis of the TP solution of (\ref{Spsist}) reveals that at this bifurcation a pair of complex conjugate eigenvalues cross the imaginary axis, destabilizing the TP in an oscillatory manner.
The corresponding eigenfunction is localized and has its maximum modulus at the back of the pulse.

Numerical simulations of a TP for parameters beyond this bifurcation show that for values of $\mu_2$ in the lower part of Fig.~\ref{fig:stability} the Hopf is supercritical and the system tends to a stable breathing traveling pulse-like solution like the one shown in Fig.~\ref{fig:stability}I. This pulse propagates while presenting low amplitude periodic oscillation of its width and amplitude, specially on the back of the pulse. The amplitude of the oscillations increases as the parameters are moved away from Hopf bifurcation. These results are similar to those shown in \cite{Or-Guil2001}.
For larger values of $\mu_2$ the Hopf is subcritical and initial conditions close to the unstable TP end up in failure of propagation, as shown on Fig.~\ref{fig:stability}II. The TP initially shows some oscillations, but after propagating for some time its amplitude decreases until vanishing eventually, decaying to the rest state.

%-----------------Subsec-------------%
\subsection{SNIC bifurcation}
\label{subsec:snicTP}

The stability region of TPs is limited on the right by the $SNIC$ bifurcation (dot-dashed purple line). The TP exists all the way until the $SNIC$ line, experiencing a divergence of its width as the parameters approach the $SNIC$ \cite{arinyoiprats2021traveling}. More details of such bifurcation are given in Sec. \ref{SNSL_MSDS}.
After crossing the $SNIC$ bifurcation line in parameter space the $P_1$ homogeneous steady solution disappears when colliding with the $P_2$ fixed point in a saddle-node, leading to a homogeneous oscillation.
A TP like initial condition for these parameter values forms a pulse on top of the homogeneous oscillation (see Fig. \ref{fig:stability}VI). 
As the background oscillates, the localized structure corresponding to a TP before the bifurcation is now reset at each oscillation. After propagating briefly, the background oscillation brings the localized structure back to the initial position. 
The system shows then an almost periodic behavior, as it can be seen in Figure~\ref{fig:stability}VI.
For some of the simulations we did for different parameter values we observed a drift in the position of the localized structure after each oscillation of the ground state. This drift is much slower than the velocity of the localized structure during the oscillation.

%-----------------Subsec-------------%
\subsection{Double heteroclinic bifurcation I}
\label{subsec:hetITP}

On the lower right part of the stable region the $SNIC$ line terminates at a $SNSL$ codimension-2 point and the stability region is from there on delimited by a monotonic double heteroclinic bifurcation line $DH_{1m}$, starting from the same point, and a fold of TPs (Fold $DH_1$). The double heteroclinic bifurcation is explained in more detail in Sec. 
\ref{Heterocl_MSDS}. 
Past these bifurcations, a pulse develops a protuberance in its tail that will eventually generate two pulses  propagating in opposite directions (see Fig.~\ref{fig:stability}IV). 
This process repeats generating two wave trains propagating in opposite directions, and it is known in the literature as "backfiring" \cite{Bar1994}.

%-----------------Subsec-------------%
\subsection{Double heteroclinic bifurcation II}
\label{subsec:hetIITP}

At the upper part, the stability region is delimited by a monotonic double heteroclinic bifurcation and a fold of TPs, labeled as $DH_{2m}$ and Fold $DH_{2}$ in Fig. \ref{fig:stability}. A detail description of the scenario leading to such bifurcations is given in Sec. \ref{Heterocl_MSDS}.  An initial pulse for parameter values past the $DH_{2}$ bifurcation starts losing its shape and, after some time, the tail of the pulse grows a perturbation which, eventually, travels as a pulse in the opposite direction. Both pulses start generating new pulses that annihilate when colliding and, sometimes, generate even more pulses. This process generates a spatiotemporal chaotic regime that propagates until taking up all the system.  An example of such dynamics is shown in Figure \ref{fig:stability}V, which resembles spatio-temporal intermittency \cite{Chate1987}. This turbulent regime is also related to the phenomenon of "backfiring" \cite{Bar1994}.

%__________-----------------Section-------------____________%
\section{MSDS bifurcations}
\label{MSDS_bif}
In this section we analyze in detail some of the bifurcations discussed above from the point of view of the MSDS (Moving Spatial Dynamical System).

In the PDEs (\ref{Spsist}), TPs are codimension-0, i.e. they exist for any parameter values within the existence region, where their velocity is uniquely determined by the parameters. However, in the MSDS (\ref{Spsistrav}), where the velocity $c$ of the moving reference frame is a free parameter, TPs are codimension-1, i.e., once the other parameters are fixed, they only exist for the precise value of $c$ corresponding to the velocity of the TP in the PDEs. As a TP corresponds to a homoclinic trajectory in the MSDS, this value of $c$ indicates the exact location of a homoclinic bifurcation in the ($\mu$,$c$) parameter space. For the parameters values where $P_1$ is a Saddle Bifocus (see Appendix~\ref{local_MSDS}) the homoclinic bifurcation shows Shilnikov characteristics \cite{Ovsyannikov1992}. This implies the existence complex limit cycles in the MSDS, which describe spatially chaotic traveling structures. The Shilnikov's effect and the presence of these structures do not affect the results presented in this work.

%-----------------Subsec-------------%
\subsection{SNIC as a SNSL in the MSDS}
\label{SNSL_MSDS}
\begin{figure}
 \centering
\includegraphics[width=0.49\textwidth]{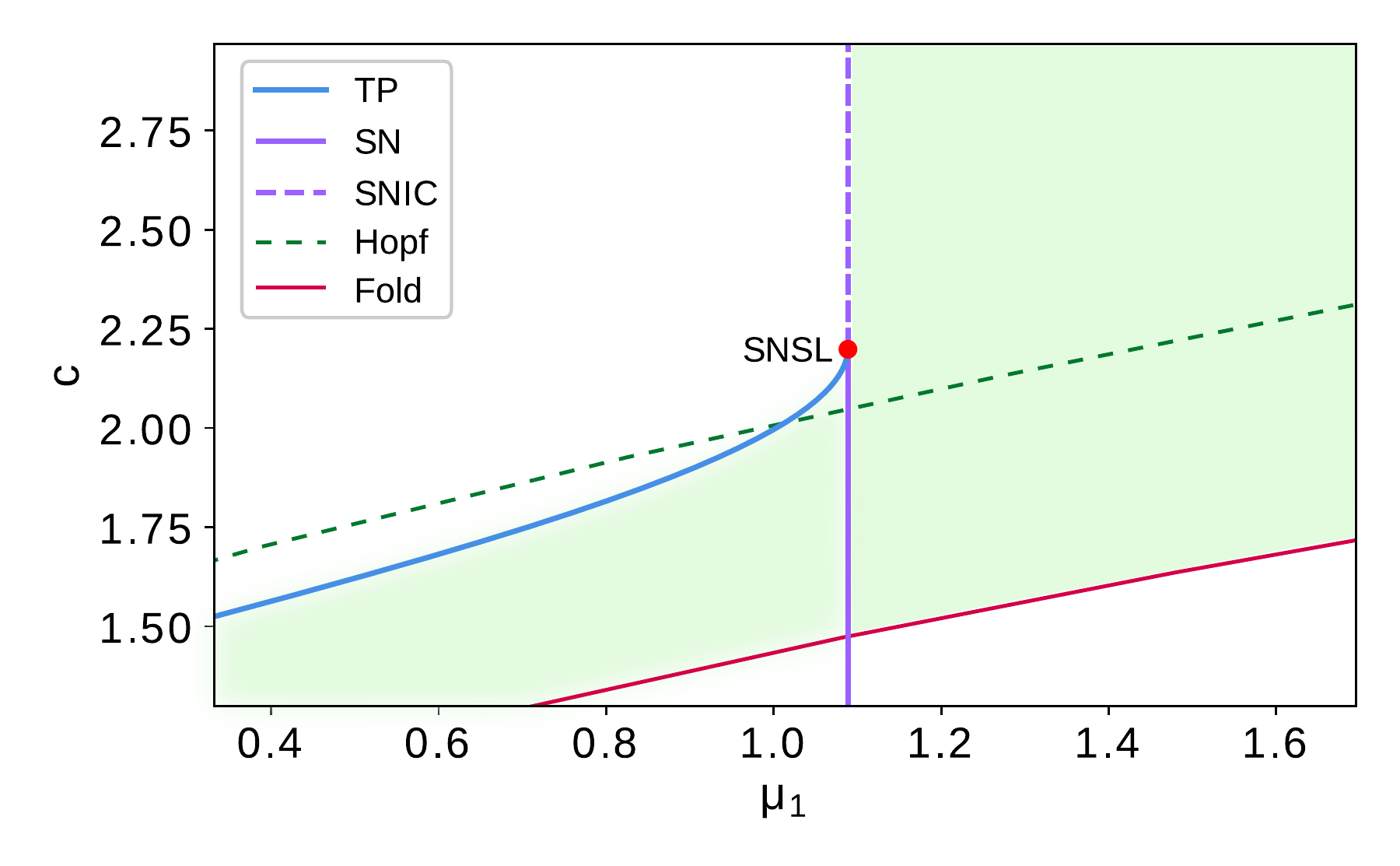}
\caption{\label{fig:SNSL_TP} Phase diagram of the MSDS for $\mu_2=2$, close to the $SNIC$ bifurcation of the temporal system (indicated by the vertical purple line). The blue line is the homoclinic bifurcation whose associated homoclinic trajectory corresponds to a TP. This bifurcation line ends at a codimension-2 point ($SNSL$) when colliding with the $SN$ bifurcation of the homogeneous solutions (purple line). The $SNSL$ indicates the end of the existence of TPs in correspondence with the $SNIC$ of the homogeneous solutions. The green shaded region indicates the existence of limit cycles in the MSDS, i.e. wave trains on the PDEs, whose period diverge at the $SNIC$ and homoclinic bifurcations. The dashed green and red lines indicate, for completeness, a Hopf bifurcation of $P_3$ and a Fold of Cycles where such wave trains are created or destroyed.}
\end{figure}

The $SNIC$ bifurcation of the temporal system is somehow more complicated in the MSDS.
In the MSDS the $SN$ and the homoclinic bifurcations associated with the TPs are codimension-1. The bifurcation where the TPs are destroyed is described by a codimension-2 bifurcation where these two manifolds meet.
The full unfolding of this bifurcation point is shown on Fig.~\ref{fig:SNSL_TP}.
The $SN$ of the MSDS occurs for the value of $\mu_1$ corresponding to the $SNIC$ bifurcation of the temporal system, and is indicated by the vertical purple line in Fig.~\ref{fig:SNSL_TP}. 
To the left of this line, TPs correspond exactly to the homoclinic trajectory created at the homoclinic bifurcation (blue line in Fig.~\ref{fig:SNSL_TP}). 
To the right they do not exist as explained in Section \ref{subsec:snicTP}. Then, following the homoclinic bifurcation line, we observe that TPs terminate at the precise value of $\mu_1$ where the homoclinic bifurcation line (in blue) touches tangentially the SN bifurcation (purple vertical line).This codimension-2 point corresponds to a $SNSL$ in the MSDS, marked as a red point on Fig~\ref{fig:SNSL_TP}. Therefore, the $SNIC$ bifurcation of a TP is always a $SNSL$ in the MSDS. Notice that close enough to the $SNSL$ the eigenvalues of $P_1$ are real (see Appendix~\ref{local_MSDS}) and the homoclinic bifurcation associated with the TP does not show Shilnikov's effect.

This $SNSL$ separates two different cases of the $SN$ bifurcation where $P_{1}$ and $P_{2}$ collide.
For velocities $c$ above the $SNSL$ point, the $SN$ is a saddle node on the invariant cycle bifurcation ($SNIC$, dashed purple line). 
The invariant cycle on where the homogeneous solutions appear describes (not necessarily temporally stable) wave trains, shown in Fig.~\ref{fig:SNSL_TP} as shaded green region. The period of these large velocity wave trains diverge as the parameters approach the  $SNIC$ bifurcation from the right, and, therefore, do not coexist with the TP.
For velocities below the $SNSL$, the $SN$ of the homogeneous solutions corresponds to a saddle node off the invariant cycle (solid purple line) and, therefore, (not necessarily temporally stable) slow wave trains expand inside the excitability region all the way to the homoclinic bifurcation (blue line), where their period also diverges.
The origin of these wave trains on the MSDS can be linked to a Hopf and a Fold of cycles shown in dashed green and red lines in Fig.~\ref{fig:SNSL_TP} respectively.  These bifurcations are not relevant for the main discussion of this work and are only shown for completeness. 
 
%___________________figure__________________%
%________________End figure_________________%

\begin{figure}[h!]
 \centering
\includegraphics[width=0.49\textwidth]{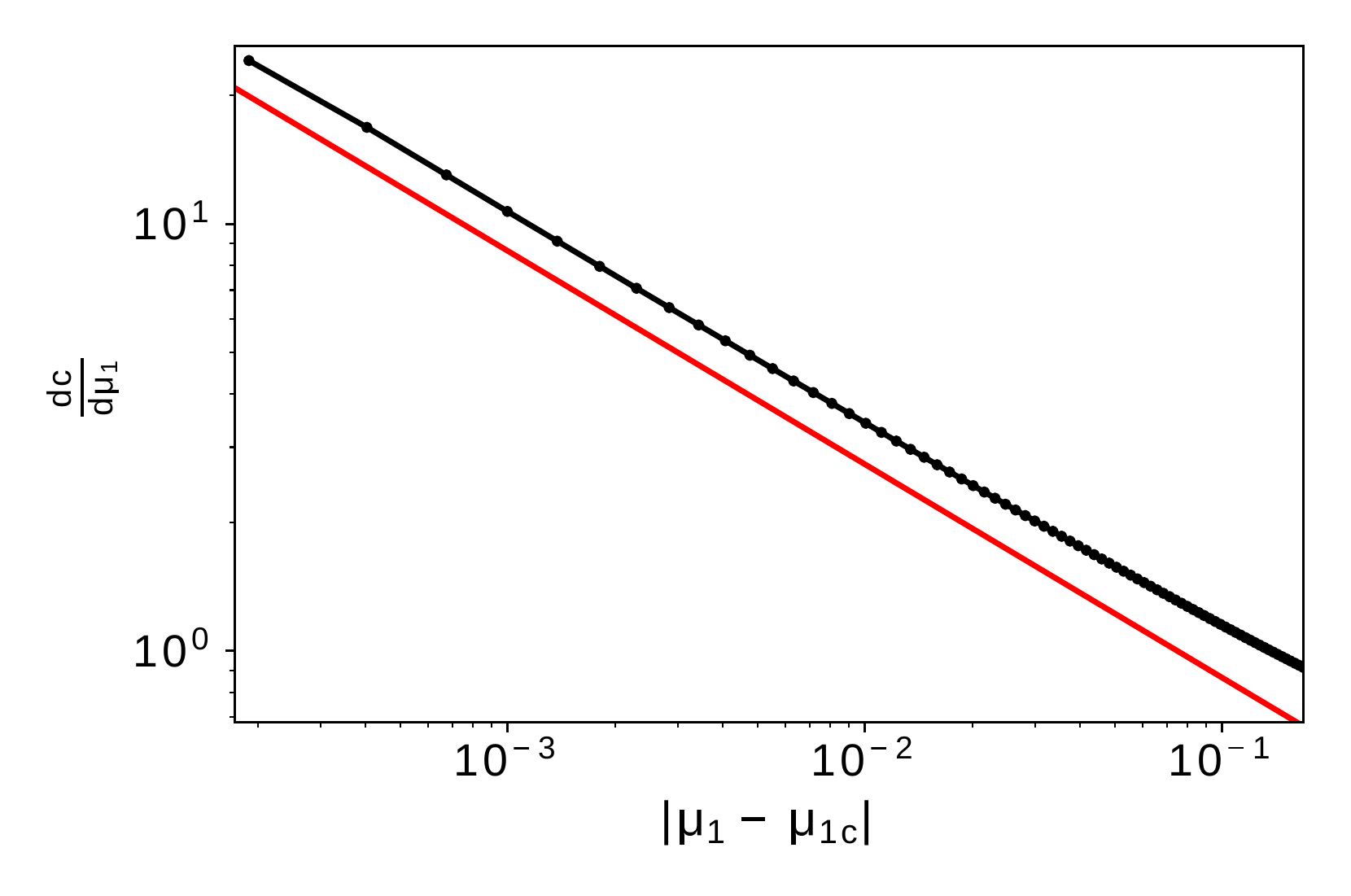}
\caption{\label{fig:SNSL_c} Scaling of the derivative of the velocity of the TP with respect to $\mu_1$ in the proximity of the $SNSL$. The expected theoretical scaling is shown in red for comparison.}
\end{figure}

The proximity to the $SNSL$ bifurcation affects the TP in two different ways. First, it slows down the approach to the rest state, being on the $SNSL$ slower than exponential. This effect was studied in detail in \cite{arinyoiprats2021traveling} and is associated with the tendency of the slow eigenvalue of $P_1$ when approaching the $SN$. 
Second, the derivative of the pulse velocity, which is the value of $c$ of the homoclinic bifurcation, with respect to the control parameter ($\mu_1$) diverges when approaching the bifurcation as $\frac{d c}{d\mu_1} \propto \frac{1}{\sqrt{\mu_1}}$. 
This scaling is shown in Fig. \ref{fig:SNSL_c}.

%___________________figure__________________%
%________________End figure_________________%

\subsection{Double heteroclinic bifurcations I and II}
\label{Heterocl_MSDS}
%___________________figure__________________%
 \begin{figure}[h!]
 \centering
\includegraphics[width=0.49\textwidth]{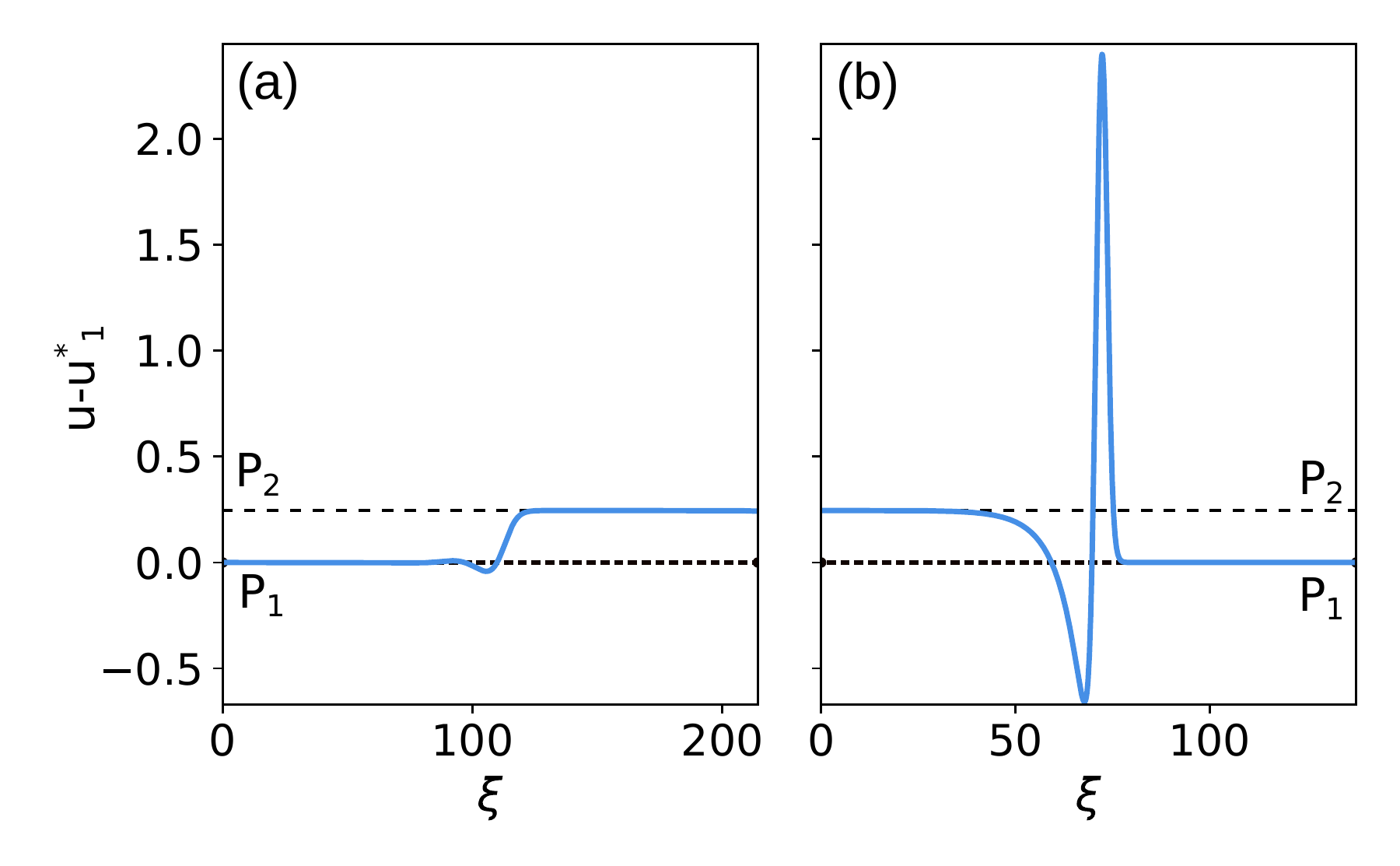}
\caption{\label{fig:Fronts} The two heteroclinics connections (fronts in the PDE) between $P_1$ and $P_2$ on the $DH_{2m}$ point for $\mu_1=0.4$. a) Back-heteroclinic connection ($h_b)$. $h_b$ is part of a parametrized by velocity family of fronts of $P_1$ into $P_2$, its velocity is selected by the velocity of $h_f$. b) Front-heteroclinic connection ($h_f$).}
\end{figure}
%________________End figure_________________%

The double heteroclinic bifurcations I and II described in Sections \ref{subsec:hetITP} and \ref{subsec:hetIITP} are associated with double heteroclinic connections between $P_1$ and $P_2$ in the MSDS, where each heteroclinic describes a different propagating front in the PDEs \cite{Sandstede2000}. 
The first of these heteroclines, which we name back-heteroclinic connection ($h_b$), is given by a transverse intersection of the 2-dimensional unstable manifold of $P_1$ and the 3-dimensional stable manifold of $P_2$.
The $h_b$ is then a codimension-0 solution of the MSDS.
This implies the existence of a continuous family of propagating fronts of $P_1$ into $P_2$ for parameter values in the neighborhood of the bifurcation point, which are characterized by their velocity. 
An example of this front is shown on Fig.~\ref{fig:Fronts}a.

The other heteroclinic connection, which we name front-heteroclinic connection ($h_f$), is given by the tangential intersection of the  1-dimensional unstable manifold of $P_2$ and the 2-dimensional stable manifold of $P_1$.
This connection is codimension-2 in the MSDS, which implies that the propagating front of $P_2$ into $P_1$ is not generic and exists only at the bifurcation.
An example of this front is shown on Fig.\ref{fig:Fronts}b.
From the interaction between these two fronts emerges the homoclinic connection corresponding to the TP. 

This double heteroclinic bifurcation ($DH$) is known in the literature as $T$ point \cite{Glendinning1986}.
Close to the $DH$ point, we can interpret the emerging traveling pulse as the pinning of two weakly interacting fronts, with $h_b$ chasing $h_f$ separate by a large plateau close to $P_2$. The plateau width will diverge to infinite while approaching the $DH$ point. 

From the $DH$ point, another homoclinic connection emerges, in this case biasymptotic to the $P_2$ solution, associated with an unstable traveling pulse on top of the \textbf{$P_2$} homogeneous solution. 
Similar to the traveling pulse on $P_1$, these pulses will generate a plateau, in this case around $P_1$ as the parameters approach the $DH$ point. Close to the bifurcation these traveling pulses can be interpreted as two weakly interacting fronts, in this case with $h_f$ chasing $h_b$, separated by a large plateau close to $P_1$. This plateau diverges to infinite as the parameters approach the $DH$ point. We will refer to these traveling pulses with $P_2$ as background as TP$_2$.

Strictly speaking, the $DH$ point exists only for infinite-domain systems. In finite systems with periodic boundary conditions we find, instead, a transition from the TP with $P_1$ background to the traveling pulse with $P_2$ background (TP$_2$). This transition occurs in one (or more \cite{Or-Guil2001}) folds of the traveling pulses close to the region in the parameter space where the $DH$ point is found in the infinite-domain system.

The weak interaction between fronts is given by the overlap of their asymptotic decay to $P_2$ \cite{Coullet2002,Gomila2015}.
This decay is determined by the spatial eigenvalues of $P_2$.
Two different scenarios can be found on this $DH$ point depending on how $h_b$ tends asymptotically to $P_2$:

\subsubsection{Monotonic $DH$ point}
\begin{adjustwidth}{0.5cm}{0pt}

We first discuss the case when $h_b$ tends to $P_2$ monotonically, i.e. when the stable spatial eigenvalues of $P_2$ are real at the $DH$ point.
In this case, the interaction between a $h_b$ and a $h_f$ fronts separated by a plateau on $P_2$ decreases monotonically with the distance between the fronts \cite{Coullet2002}. As a consequence, the TP branch approaches the $DH$ point also monotonically, without snaking (Fig. \ref{fig:T_monotonic}).

%___________________figure__________________%
 \begin{figure}
 \centering
\includegraphics[width=0.49\textwidth]{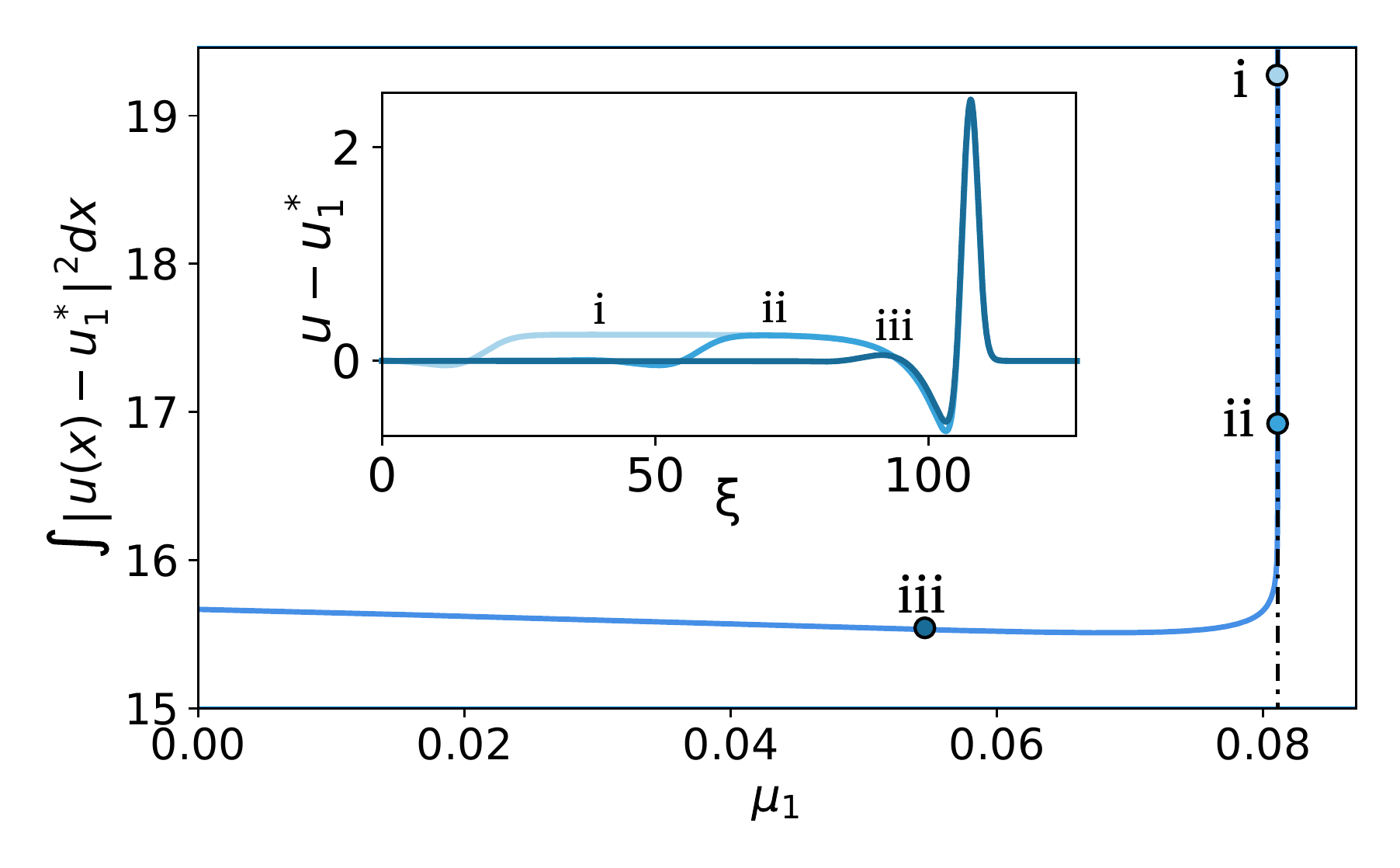}
\caption{\label{fig:T_monotonic} Bifurcation diagram of the TP solution close to the $DH_{1}$ point for $\mu_2=0.4$ (marked with a vertical dashed dotted line). The width of the TP, as measured by the norm of the TP, diverges monotonically at the bifurcation. Inset shows the profile of the TP for the three different values of $\mu_1$ indicated in the main figure.}
\end{figure}
%________________End figure_________________%

As mentioned above, when approaching the $DH$ point the TP starts to present a plateau around $P_2$ (see the inset of Fig. \ref{fig:T_monotonic}). The plateau width diverges logarithmically with the distance of the control parameter to the $DH$ point \citep{arinyoiprats2021traveling}.

Close to the $DH$ point, where the TP has an arbitrarily long plateau tending to the unstable homogeneous solution $P_2$, one could expect the TP to be unstable, but, actually, in this case the plateau is only convectively unstable in the co-moving reference frame, being the TP globally stable. The other case, where the plateau is absolutely unstable, corresponds to the collapsed snaking case we will discuss below \cite{Or-Guil2001,Sandstede2000,Nii2000}.
Close enough to $SN_{r-}$ the $DH$ point is always monotonic.
\end{adjustwidth}

\subsubsection{Collapsed snaking}
\begin{figure}
 \centering
\includegraphics[width=0.49\textwidth]{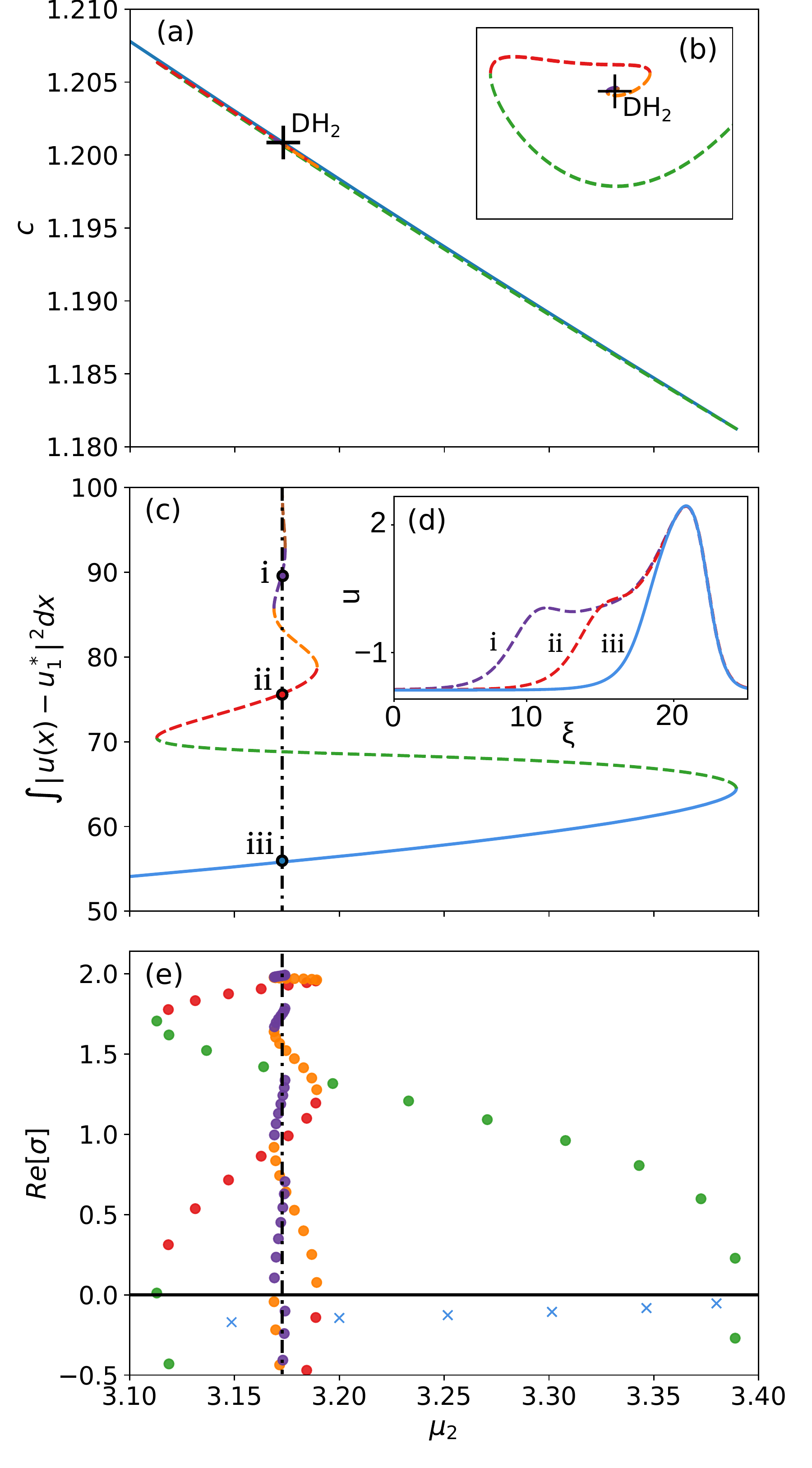}
\caption{\label{fig:T_Sneaking} \textbf{a)} Phase diagram of the MSDS in the $(c,\mu_2)$ parameter space for $\mu_1=0$ and around the $DH_2$  bifurcation point (indicated by a black cross). The panel can also be interpreted as a bifurcation diagram of the PDEs where the curves represent different branches of TPs, spiraling around $DH_2$. The axis orientation makes difficult to differentiate the different lines, so in panel \textbf{b)} we have stretched the parameter space around $DH_2$ to better show the spiraling of the bifurcation lines. The axis of this inset are no longer $\mu_2$ and c, but a linear combination of them. \textbf{c)} Bifurcation diagram of the TP solution close to the $DH_{2}$ point in the Collapsed Snaking case for $\mu_1 = 0$. Solid (dashed) lines indicate stable (unstable) solutions. The value of $\mu_2$ corresponding to the $DH_{2}$ point is marked with a vertical dashed dotted black line. \textbf{d)} Spatial profiles of the field $u(\xi)$ for the different branches indicated in the main figure. \textbf{e)} Real part of the most unstable eigenvalues of the TP as function of the parameter $\mu_2$. Dots represent real eigenvalues and crosses complex conjugate pairs. The color indicates the corresponding branch of the TP. Translationally invariant zero eigenvalues have been omitted.}
\end{figure}

\begin{adjustwidth}{0.5cm}{0pt}

In this subsection, we consider the case when the asymptotic tendency of $h_b$ to $P_2$ is oscillatory, i.e. when $P_2$ has complex eigenvalues at the $DH$ point.
In this case, the interaction between $h_b$ and $h_f$ changes periodically from attractive to repulsive with the distance between them, while the interaction strength decays exponentially, allowing two fronts to lock (or pin) at discrete separation lengths.
This makes the homoclinic bifurcation curve corresponding to the TP to snake towards the $DH$ point (Fig. \ref{fig:T_Sneaking}).
From the PDE point of view, the TP has to turn infinitely many folds while the parameters approach the $DH$ point. At each fold, the width of the pulse increases in half the wavelength of the asymptotic oscillations of the front $h_b$ profile. As a result of the fronts locking, at the $DH$ point there are infinitely many TPs with different widths. The envelope of the bifurcation line shows a characteristic divergence as a function of the distance to the $DH$ point as in the previous case. This phenomenon is known as collapsed snaking \cite{Burke2007}.
\end{adjustwidth}

%___________________figure__________________%
%________________End figure_________________%

\begin{adjustwidth}{0.5cm}{0pt}
In this case, the TP loses its stability when the plateau starts to form after the first fold, and at each successive fold it gets an extra positive eigenvalue (Fig.~\ref{fig:T_Sneaking}~e) \cite{Or-Guil2001}. This is different from the usual collapsed snaking, where the stability changes at each fold, the reason being that, in this case, the homogeneous state of the plateau is absolutely unstable.
\end{adjustwidth}

~\\

The approach of the TP$_2$ to the $DH$ point presents both same scenarios as the TP.
Monotonic or oscillatory is determined in this case by the eigenvalues of $P_1$ at the $DH$ point. The approach is monotonic if the eigenvalues of $P_1$ are real, and snaking is observed otherwise. In particular, the approach is monotonic close enough to $SN_{r-}$.

The transition between both scenarios is given by a codimension-3 point (codimension-2 in the PDEs parameter space) given by the transverse intersection between the $DH$ bifurcation codimension-2 manifold and a Belyakov-Devaney pseudo-bifurcation codimension-1 manifold, (see Appendix~\ref{local_MSDS}).
At this codimension-2 point, the double heteroclinic connection occurs while the $P_2$ point have an algebraically degenerate real eigenvalue.
The Belyalok-Devaney pseudo-bifurcation indicates, precisely, the drift speed at which the $P_2$ homogeneos solution changes from being convectively unstable to being absolutely unstable. Further information about convective and absolute instabilities can be found in \cite{walgraef2012spatio,Sandstede2000}.

\subsection{Connection between temporal and spatio-temporal bifurcations of TPs.}

%___________________figure__________________%
 \begin{figure}
 \centering
\includegraphics[width=0.49\textwidth]{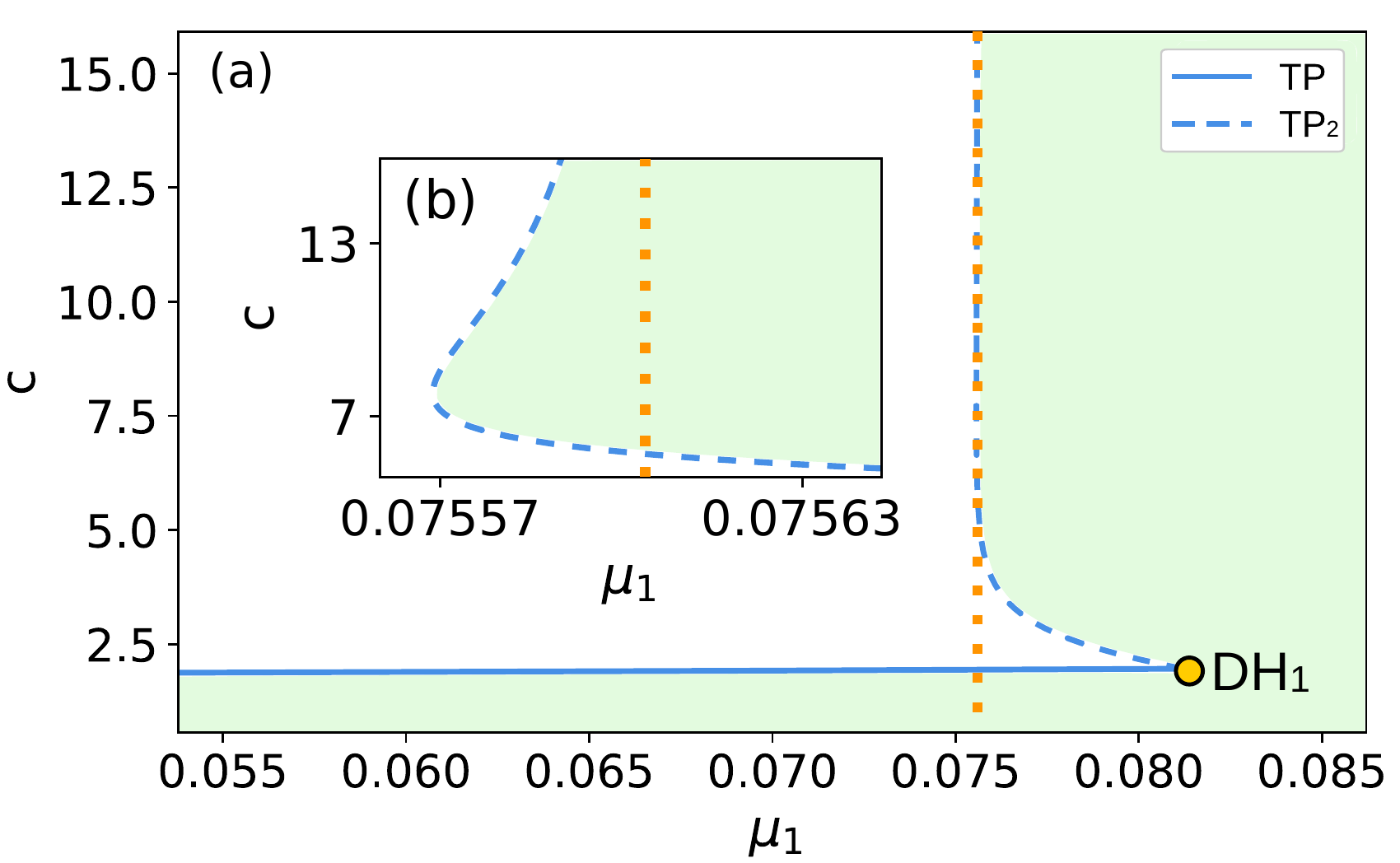}
\caption{\label{fig:T-Homo} \textbf{a)} Bifurcation diagram around the $DH_{1}$  bifurcation point for $\mu_2=0.4$. The homoclinic bifurcations creating the 
TP (solid blue) and TP$_2$ (dashed blue) converge at the $DH$ point. The velocity of the TP$_2$ has an asymptotic divergence when approaching the temporal homoclinic bifurcation point ($L^{b}_-$), indicated by the vertical dotted orange line. At this point and in the $c\rightarrow \infty$ limit the TP$_2$ converges to the homogeneous temporal homoclinic connection.\textbf{b)} Detail of $TP_2$ close the parameter value of $L^{b}_-$.}
\end{figure}
%________________End figure_________________%

In this section, we discuss the connection between the (spatial) bifurcations of the MSDS and the (temporal) bifurcations of the local dynamics.
We focus on the homoclinic bifurcation  associated to the TP$_2$ in the limit $c\rightarrow \infty$. In this limit, as already mentioned in Sec. \ref{MSDS}, the MSDS recovers the temporal equations.
In particular the TP$_2$ emerges from the temporal homoclinic bifurcation for $c=\infty$ and becomes a stable excitable TP at the $DH_{1}$  point for a finite velocity.

The homoclinic connection that forms at the homoclinic bifurcation unfolding from $DH_{1}$  ($DH_{2}$) (dashed line in Fig. \ref{fig:T-Homo}) corresponds to TP$_2$. 
We observe how, when the parameters approach the values associated with the homoclinic temporal bifurcation $L_-^{b}$ ($L_-^{s}$), (dotted vertical line) the velocity of the TP$_2$ diverges to infinity.
At the same time, the dynamics of the MSDS slows down making the pulse wider.
These two asymptotic behaviors show the tendency of the solution of the MSDS in the limit $c\rightarrow \infty$, where the local dynamics of the system are recovered.
Therefore, the homoclinic connection that TP$_2$ represents in the MSDS tends in this limit to the temporal homoclinic connection of the local dynamics of the system on $L_-^{b}$ ($L_-^{s}$). 

This connection between the TP$_2$ and the homoclinic trajectory of the temporal system establishes a bridge between the temporal excitable trajectory and the traveling pulses \cite{arinyoiprats2021traveling}.

\begin{figure}
 \centering
\includegraphics[width=0.49\textwidth]{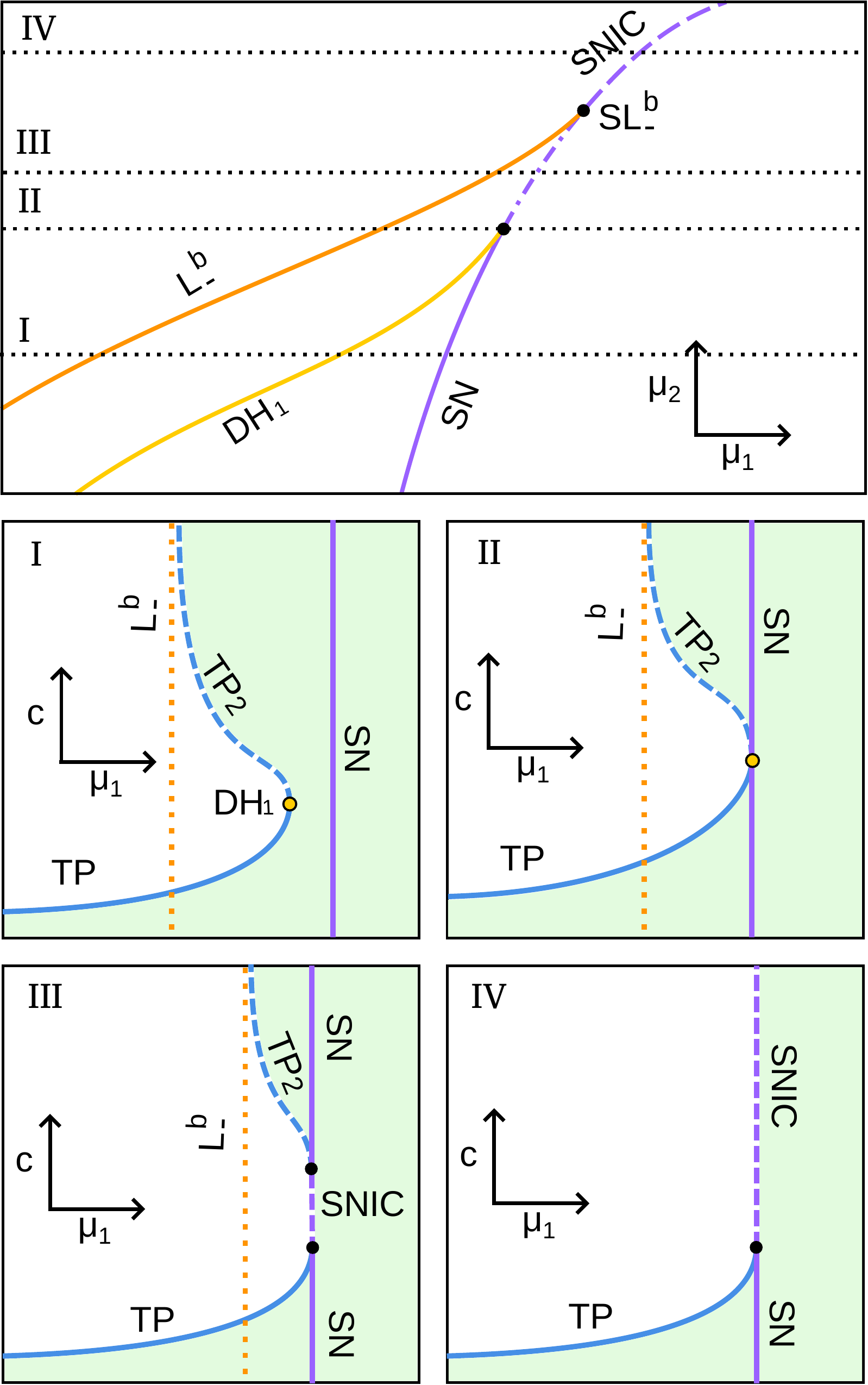}
\caption{\label{fig:SP_connect} Schematic phase diagram around the $SNSL_-^{b}$ which shows the tangential point between the temporal bifurcation line $L_-^{b}$ and the $SN_r$; and the spatio-temporal bifurcation $DH_1$ and the $SN_r$. In the four small down panels are shown bifurcation diagram of the PDEs (related with phase diagrams of the MSDS) for the four different fixed $\mu_2$ lines shown in the phase diagram. On this bifurcation diagram the solid blue lines represent the TP with rest point on $P_1$, the dashed blue lines represent the unstable TP with rest state on $P_2$, the purple line represent the $SN$ bifurcation on the MSDS and the orange line shown the $\mu_1$ value where the $L_-^{b}$ occurs in the temporal model. Green colored region of the bifurcation diagrams represent traveling waves (oscillatory region of the MSDS).}
\end{figure}

Furthermore, the connection between $DH_{1}$  and $L_-^{b}$ allows us to understand the transition between $DH_{1}$  and $SNSL$ of TPs bifurcations.
This transition is illustrated in Fig.~\ref{fig:SP_connect}, where different cuts of the parameter space of the PDEs are shown.
The panels I to IV, corresponding to the four different cuts of the main figure indicated by the horizontal dotted lines,  can be interpreted as bifurcation diagrams of the PDEs or phase diagrams of the MSDS.

Close enough to the $SN_{r-}$ bifurcation point both, $P_1$ and $P_2$, have real eigenvalues and therefore the approach of the TP and the TP$_2$ to the $DH_1$ point is monotonic.
This is illustrated in a first cut of the parameter space, where the $DH$ bifurcation occurs between the $L_-^{b}$ and the $SN$ (Fig.~\ref{fig:SP_connect}-I).
Notice the similarities between this schematic representation and the numerical result shown in Fig.~\ref{fig:T-Homo}.

The $DH$ point approaches the $SN$ when increasing $\mu_2$ until it tangentially touches the $SN_r$ bifurcation.
At this codimension-2 point of the PDEs (codimension-3 in the MSDS) the system presents a homoclinic connection of a non-hyperbolic point.
A bifurcation diagram cutting through this high codimension point is shown in Fig.~\ref{fig:SP_connect}-II.

Beyond this point the TP and TP$_2$ branches are separate, ending each one at the $SN$ in a $SNSL$ of the MSDS.
The $SN$ bifurcation between these two $SNSL$ is a $SNIC$ of the MSDS.
This case is illustrated in Fig.~\ref{fig:SP_connect}-III.

Eventually, the $L_-^{b}$ bifurcation touches the $SN_r$ bifurcation, corresponding with the $SNSL$ of the temporal system.
Beyond this point the TP$_2$ does not exist anymore, while the TP still ends in a $SNSL$ bifurcation of the MSDS. 
This case correspond to the one described in Sec.~\ref{SNSL_MSDS}.
A schematic phase diagram of this scenario is illustrated in Fig.~\ref{fig:SP_connect}-IV.
Notice the similarities between this scheme and the numerical result shown on Fig.~\ref{fig:SNSL_TP}.

%___________________figure__________________%
%________________End figure_________________%

Other bifurcations of the temporal system percolate in the MSDS through the limit $c\rightarrow \infty$.
This is the case of the temporal Hopf bifurcation and Fold of cycles, which in the MSDS generate the cycles (waves trains in the PDEs).
Higher codimension points involving these bifurcations, such as degenerate loops points, could affect the stability of the pulses.
These bifurcations have not been observed for the considered parameter values.
Nevertheless, it is expected that for other values of the parameters these bifurcations will have an effect on the stability of the pulses.

\section{Drift Pitchfork}
\label{DP}

The existence of TPs can be tracked down to the $SN_r$ for $c=0$, in particular for the same parameters of $SN_{r-}$ in the temporal case.
For $c=0$ the fixed points belong to the symmetry plane $\Pi$ and therefore the linearization around them is strongly symmetric.
In particular the eigenvalue problem on $SN_{r-}$ is degenerate, having the degenerate fixed point two opposite real and two non-diagonalizable zero eigenvalues. 
This point is known as Reversible Takens-Bogdanov ({\it RTB}).

\begin{figure}
 \centering
\includegraphics[width=0.49\textwidth]{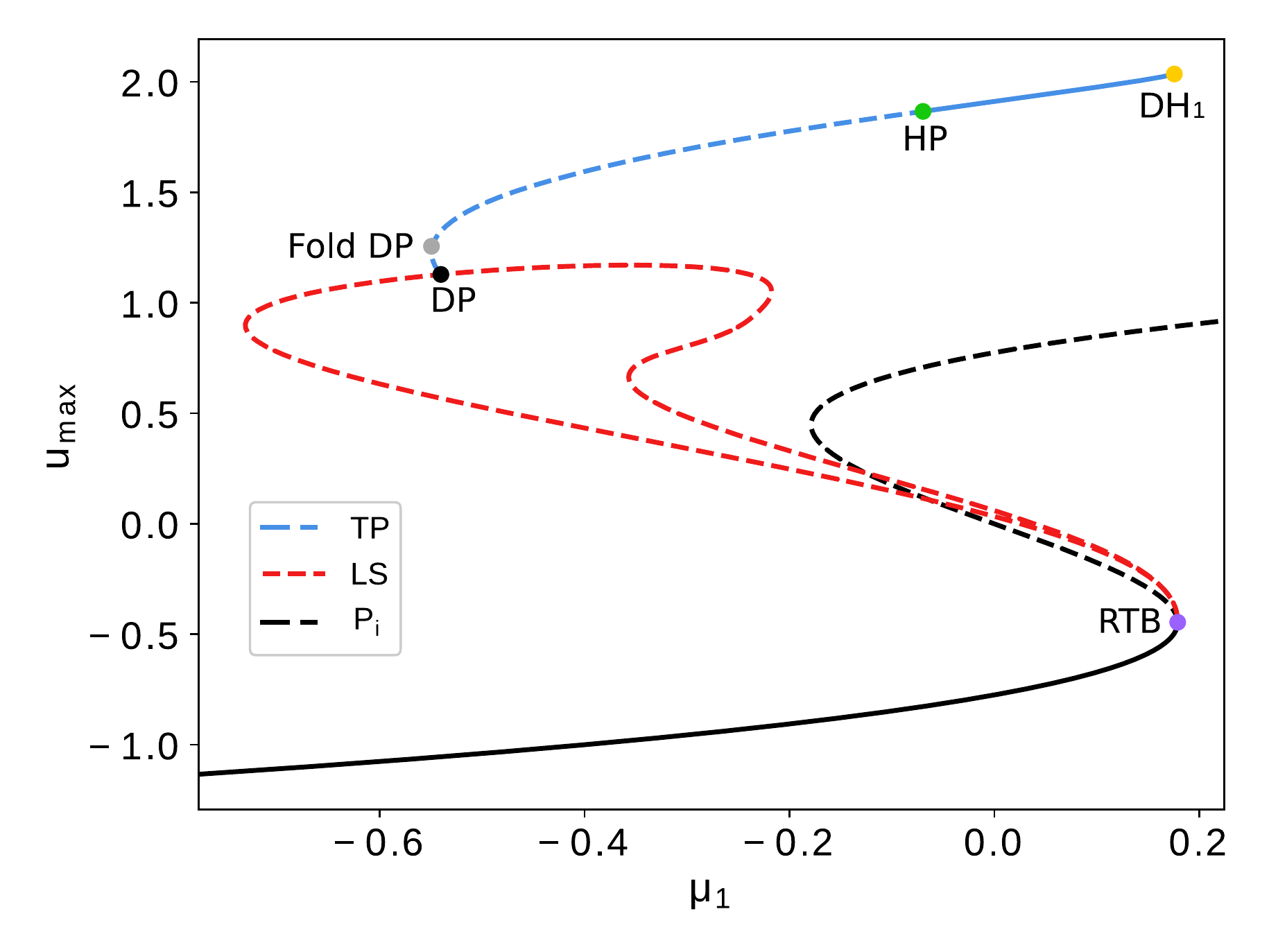}
\caption{\label{fig:DP} Bifurcation diagram of LS and TP solutions for $\mu_2 = 0.6$. The diagram shows the maximum value of u ($u_{max}$) of the different solutions as function of $\mu_1$. Solid (dashed) lines represent stable (unstable) solutions. The black line represents the homogeneous solutions, the red line the LS, and the blue line the TP. The LS branch starts from the right $SN$ ($SN_r$, {\it RTB}) represented by a purple dot. The TP branch starts at the Nonequilibrium Ising-Bloch transition, also known and Drift-Pitchfork ($DP$) shown as a black dot, followed by its associated fold of pulses showed as a gray dot. The TP changes its stability at the Hopf bifurcation ($HP$) (green dot) and ends at the $DH_1$  bifurcation (light orange dot).}
\end{figure}

A small amplitude homoclinic reversible connection of the fixed point $P_1$ is part of the unfolding around the {\it RTB} \cite{Champneys1998,Haragus2011}. 
This homoclinic connection is translated into low-amplitude achiral steady localized structures (LS) of the PDEs.
This LS is unstable, and its amplitude grows decreasing $\mu_1$ away from the {\it RTB} until it reaches a fold (Fig.~\ref{fig:DP}).
After the fold, it keeps growing in amplitude, but now increasing $\mu_1$, until it reaches a second fold. After this second fold the LS starts changing its shape: the middle point of the pulse starts developing a local minimum for both ($u,v$) fields and the pulse starts decreasing in amplitude.
After a third fold, while it reaches out again to the {\it RTB}, the pulse decreases in amplitude while the local minimum in the middle of the pulse approaches $P_1$, effectively splitting the LS in two.
This approach to the fixed point comes with a slowing down in the $\xi$ dynamics and, close enough to the {\it RTB}, the two small amplitude peaks separate half of the system size.
From this endpoint starts a new branch of LS very similar to the previously discussed one, but with two LS separated half the system size. Similar behavior has been found in other systems \cite{Ruiz-Reynes2020a,Yochelis2008}.

TPs are generated from this LS at a Nonequilibrium Ising-Bloch transition, also known as Drift-Pitchfork bifurcation ($DP$), between the first and second folds of the LS \cite{Michaelis2001,Tlidi2009,Paulau2009}.
The emerging TP breaks the chiral symmetry of the LS and propagates in space with non-zero velocity $c$. The chiral forks of the pitchfork represent the two mirror-image TP propagating in opposite directions. Some vegetation models shows similar mechanism for creation of traveling pulses outside the excitable region \cite{Marasco2014,Iuorio2020}.

%___________________figure__________________%
%________________End figure_________________%

The $DP$ bifurcation changes from supercritical to subcritical, on a degenerate codimension-2 point.
A fold of traveling pulses is part of the unfolding of this high codimension point.
This fold is shown as a gray line on Fig. \ref{fig:stability} and as a gray dot in Fig.~\ref{fig:DP}.

\section{Conclusion}
\label{Conclusions}

We have studied the mechanisms behind the creation (or destruction) of traveling pulses in a general model for Type-I excitable media. 
Traveling pulses are stable only in part of the excitable region of the local dynamics. 
They destabilize through different transitions related with the bifurcations of the local dynamics delimiting the excitable region in the parameter space, but  can also undergo purely spatiotemporal instabilities. The latter, referring to Hopf bifurcation of the TP, is a secondary bifurcation of the TP leading to the destabilization of the pulses and, in some cases, the creation of other propagating localized structures (see Fig.~\ref{fig:stability} I and II).
The former, which are those that give rise to TP, i.e. the Saddle-node Separatrix-loop and the Double Heteroclinic (T point), have been studied in a comoving spatial dynamics description (Moving Spatial Dynamical System, MSDS), that has been shown to converge to the temporal dynamics in the limit in which the speed is very large.
Exploiting this limit we have been able to connect these bifurcations of the traveling pulses to the temporal bifurcations leading to Type-I excitability, i.e., the SNIC and homoclinic bifurcations. 
Beyond this connection, we have also shown that traveling pulses bifurcate from generic steady localized structures in bistable systems through a drift pitchfork bifurcation. 

The connection between spatiotemporal and pure temporal bifurcations is general and give some insight in the properties of traveling pulses in Type-I excitable media as compared to those in Type-II. The distinction between these two case is not a simple academic exercise but has important implications when trying to model excitable behavior found in natural systems. Often the straightforward approach is using variants of a textbook model: the FitzHugh-Nagumo equation \cite{Mikhailovbook}, which is a paradigm for Type-II excitability. Instead, many models based on biological mechanisms and general principles put forward in recent years show signs of Type-I excitability, as they exhibit homoclinic phenomena, well defined thresholds, and the resultant signature on the (unbounded) periods of excitations \cite{Romeo2003,ruiz2017fairy,zhao2021,Ruiz-Reynes2022,Oto2019}. The consequence is that the observed excitable behavior might not correspond to the characteristic behavior of a Type-II excitable medium but, instead, may correspond to a Type-I excitable medium. In this work and the companion paper \cite{arinyoiprats2021traveling} we have analyzed in detail pulses that propagate in Type-I excitable media,  and their main features. We hope this will appeal to experimentalists that might be observing these novel manifestations of excitable dynamics.

Our study has described TP in an excitability scenario where the fixed points emerge from two saddle-node bifurcations organized by a cusp codimension$-2$ point. Nevertheless, some vegetation models present a variant of our route to Type-I excitable behavior based on a transcritical bifurcation, due to the constraint that populations cannot become negative \cite{Oto2019,zhao2021,Ruiz-Reynes2022}. Therefore, a natural extension of our work is to study in detail the scenario in which a transcritical bifurcation instead of a saddle-node organizes the system in parameter space. Also, in a model defined on a $1$-dimensional spatial domain, as used in the present work, it is not possible to study spatio-temporal structures like rings or spirals, that have been observed experimentally \cite{Oto2019,zhao2021}, and a logical extension is to study the scenario in a $2$-dimensional system.

We acknowledge financial support through grants RTI2018-095441-B-C22 and MDM-2017-0711 funded by MCIN/AEI/10.13039/501100011033 and by ERDF A way of making Europe. D.R-R. is supported by the Ministry of Universities through the "Pla de Recuperaci\'o, Transformaci\'o i Resil\`encia" and by the EU (NextGenerationEU), together with the Universitat de les Illes Balears. A.A-i-P. is financed by Czech Science Foundation, Project No.GA19-16066S.

\appendix
\section{Glossary of frequently used acronyms and symbols}
$P_i$: Homogeneous solutions of the Ep.~\ref{model}.

TP: Traveling pulse on $P_1$.

TP$_2$: Traveling pulse on $P_2$.

$SNIC$: Saddle-Node on Invariant Cycle.

PDEs: Partial differential equation system. Usually refer to the one given by Eq.~\ref{model}.

$SN_{l/r}$: Left/right branch of Saddle-Node bifurcation. Bifurcation of the local dynamical system.

$\lambda_{\pm}$: Eigenvalues of the local dynamical system's fixed points (Eq.~\ref{Localsist}).

$H_{1/2}$: Andronov-Hopf bifurcation of the fixed point $P_{1/2}$. Bifurcation of the local dynamical system.

$B$: Bautin point, a.k.a. degenerate Hopf. Bifurcation of the local dynamical system.

$DL^{b/s}$: Big/small degenerate saddle-loop point. Bifurcation of the local dynamical system.

$L^{b/s}_{+/-}$: Big/small saddle-loop (a.k.a. Homoclinic) bifurcation with positive/negative saddle index. Bifurcation of the local dynamical system.

$TB_{l/r}$: Takens-Bogdanov bifurcation point, located on the left/right branch of the saddle-nodes bifurcation. Bifurcation of the local dynamical system.

MSDS: Moving spatial dynamical system, given by Eq.~\ref{Spsistrav}.

$R$: Space reflection involution given by Eq.~\ref{involution}. Symmetry of the MSDS.

$\Pi$: Invariant plane under involution $R$.

$HP$: Andronov-Hopf bifurcation of TPs. Bifurcation of the PDEs.

$DH_{m/s}$: Double heteroclinic bifurcation, a.k.a. T point, monotonic/snaking. Bifurcation of the MSDS.

$DP$: Drift pitchfork bifurcation of pulses.  Bifurcation of the PDEs.

$SNSL$: Saddle-Node Separatrix-Loop of the MSDS system. Bifurcation of the MSDS.

{\it RTB}: Reversible Takens-Bogdanov bifurcation. Bifurcation of the MSDS.

\section{Fixed points and local bifurcations at the MSDS}
\label{local_MSDS}

In this appendix we study the linear dynamics around the fixed points of the MSDS. This provides the local bifurcation of the MSDS as well as the behavior of the trajectories close to the fixed points. Since the pulses and localized structures are described as homoclinics of a fixed point, the linear regime determines the shape of their tail as well as being involved in the creation and bifurcations of these structures.

Linearization of Eq.\ref{Spsistrav} around the fixed points is determined by four spatial eigenvalues given by:
\begin{align}
    \lambda'_j(P_i) =\lambda'_j(u^*_i) = \frac{-c\pm\sqrt{c^2-4\lambda_{\pm} \left( u^*_i\right)}} {2} \qquad j = 0,1,2,3
\end{align}
where $\lambda_{\pm} \left( u^*_i\right)$ are the temporal eigenvalues of the fixed point given by Eq.~\ref{eq:temp_eig}.

Firstly, we will present the codimension-0 eigenvalues configuration of the different fixed points for $c > 0$ and the codimension-1 transitions between them. Secondly, we will present the codimension-1 solutions for the steady state ($c=0$), which gives the transition between positive and negative velocity regions, and the codimension-2 transitions between them. Due to the symmetry under involution (\ref{involution}), opposite velocity will give sign changed eigenvalues.

The point $P_2$ presents two different codimension-0 configuration depending on the velocity:
\begin{itemize}
    \item $c^2>4\lambda_+$: Three real negative and a single positive eigenvalue (Saddle).
    \item $c^2<4\lambda_+$: Two complex conjugate eigenvalues with negative real part, a negative real eigenvalue and a positive eigenvalue (Saddle-Focus).
\end{itemize}

The transition between Saddle and Saddle-Focus occurs in a Belyakov-Devaney ($BD$) pseudo-bifurcation, when $c^2=4\lambda_+$.
This point has a real negative eigenvalue and two positive eigenvalues, one of them degenerated.

Points $P_{0,1,3}$ can present different eigenvalues configurations depending on the parameters and velocity:

\begin{itemize}
    \item When $\tau (u^*_i) < 0$ and $4\Delta(u^*_i) < \tau^2 (u^*_i) $ ($P_i$ is a stable node in the temporal system) the fixed point presents four real eigenvalues, two of them positive and the other ones negatives (Bisaddle).
    \item When $\tau (u^*_i) < 0$ and $4\Delta(u^*_i) > \tau^2 (u^*_i) $ ($P_i$ is a stable focus in the temporal system) the fixed point presents two complex conjugate eigenvalues with positive real part and two complex conjugate eigenvalues with negatives real part (Saddle Bifocus).
    \item When $\tau (u^*_i) > 0$ and $4\Delta(u^*_i) < \tau^2 (u^*_i) $ ($P_i$ is an unstable node in the temporal system) the fixed point presents, depending on the velocity and parameters, three different configurations of eigenvalues:
    \begin{itemize}
        \item $c^2>4\lambda_+$: Four real negative eigenvalues (Node).
        \item $4\lambda_- < c^2< 4\lambda_+$: Two real negative eigenvalues and two complex conjugate eigenvalues with negative real part (Focus-Node).
        \item $c^2<4\lambda_-$: Four by pairs complex-conjugate eigenvalues with the same negative real part.(Aligned-Bifocus).
    \end{itemize}
    \item When $\tau (u^*_i) > 0$ and $4\Delta(u^*_i) > \tau^2 (u^*_i) $ ($P_i$ is an unstable focus in the temporal system) the fixed point presents, depending on the velocity and parameters, two different configuration of eigenvalues:
    \begin{itemize}
        \item $c<\sqrt{\frac{4\Delta(u^*_i)-\tau^2(u^*_i)}{2\tau(u^*_i)}}$: Two complex conjugate eigenvalues with positive real part and two complex conjugate eigenvalues with negatives real part (Saddle-Bifocus).
        \item $c>\sqrt{\frac{4\Delta(u^*_i)-\tau^2(u^*_i)}{2\tau(u^*_i)}}$: Four by pairs complex conjugate eigenvalues with negative real part.(Bifocus).
    \end{itemize}
\end{itemize}

The transition from Node to Focus-Node configuration occurs in a $BD$ when $c^2 = 4\lambda_+$. At this transition, the fixed point presents three real negative eigenvalues, one of them degenerated.
The transition from Focus-Node to Aligned-Bifocus occurs in a $BD$ at $c^2 = 4\lambda_-$. 
At this transition the fixed points presents two complex conjugate eigenvalue and a real eigenvalues, all of them with the same negative real part.
The transition from Aligned-Bifocus to Bifocus occurs through a Degenerate Bifocus where the fixed point has a pair of complex conjugate degenerate eigenvalues. The transition from Node to Bifocus occurs in a double $BD$ point where the eigenvalue configuration has a pair of degenerate real eigenvalues. The transition from Bifocus to Saddle-Bifocus occurs thought a Hopf bifurcation where small amplitude traveling waves are created. Finally, the transition from Saddle-Bifocus to Bisaddle occurs in a Double $BD$ transition, in this case one of the degenerated eigenvalues is negative while the other one is positive.

The points $P_{1,3}$ are generated in Saddle-Node bifurcations involving $P_2$. At this codimension-1 bifurcation there are different eigenvalues compositions of the fixed point.

The fold involving a Saddle and a Bisaddle occurs through a Saddle-Node$_+$ with two negative real eigenvalues, a positive eigenvalue and a zero eigenvalue.
The fold involving a Saddle and a Node occurs through a Saddle-Node$_-$ with tree negative real eigenvalues and a zero eigenvalue.
Finally, the fold involving a Saddle-Focus and a Focus-Node takes place through a Focus-Saddle-Node$_-$ with a real negative eigenvalue, two complex conjugate eigenvalues with negative real part and a zero eigenvalue.

Fixing $c=0$ we arrived to the steady codimension-1 configurations. Even if this region is codimension-1 in the MSDS, as we are fixing the velocity, which is not a parameter of the PDEs but a condition for selected from the different traveling solutions, the steady solutions are relevant and codimension-0 in the PDEs. As the fixed points $P_i$ belong to the symmetry plane $\Pi$ the eigenvalues have a strong symmetry and can be seen as pairs of opposite complex numbers.

The point $P_2$ presents a single eigenvalue configuration when $c=0$. Two of their eigenvalues are opposite real while the other two are opposite imaginary (Center-saddle).

The points $P_{0,1,3}$ can present 3 different configurations depending on the parameters:
\begin{itemize}
    \item When $\tau (u^*_i) > 0$ and $4\Delta(u^*_i) < \tau^2 (u^*_i) $ ($P_i$ is an unstable node in the temporal system) the fixed point has two pairs of opposite imaginary eigenvalues (Double center).
    \item When $\tau (u^*_i) < 0$ and $4\Delta(u^*_i) < \tau^2 (u^*_i) $ ($P_i$ is a stable node in the temporal system) the fixed point has two pairs of opposite real eigenvalues (Reversible Bisaddle).
    \item When  $4\Delta(u^*_i) > \tau^2 (u^*_i) $ ($P_i$ is a focus in the temporal system) the fixed point has quartet of complex eigenvalues (Reversible Saddle Bifocus).
\end{itemize}

The transition from Double center to Reversible Saddle Bifocus occurs at $4\Delta(u^*_i) = \tau^2 (u^*_i) $; $\tau (u^*_i) > 0$  in a Hamiltonian Hopf bifurcation where the fixed point presents a pair of degenerate opposite imaginary eigenvalues.

The transition from Reversible Saddle Bifocus to Reversible Bisaddle occurs at $4\Delta(u^*_i) = \tau^2 (u^*_i) $ ; $\tau (u^*_i) < 0$ in a reversible $BD$ point where the fixed point presents a degenerate pair of opposite real eigenvalues.

The fold from Reversible Bisaddle to Center Saddle occurs in a Reversible Takens-Bogdanov ({\it RTB}) where the fixed point presents a degenerate zero eigenvalues and a pair of opposite real eigenvalues.

The fold from Double center to Center saddle configuration occurs in a Reversible Takens-Bogdanov-Hopf bifurcation where the fixed point presents a degenerate zero eigenvalue and two opposite imaginary eigenvalues.

\section{Numerical methods}

The temporal integration of the system have been performed using a pseudo-spectral method, adapted from \cite{Montagne}, with periodic boundary conditions on a grid with $N=4096$ nodes, a temporal step $\Delta t = 10^{-3} \ tu$, and a spatial step $\Delta x = 0.12 \ su$. The simulation shown in Fig.~\ref{fig:gauss_creation} has been initiated with a gaussian on the steady $P_1$ solution with norm $A=3$ and variance $\sigma = 0.3$. We initiate the simulation shown in Fig.~\ref{fig:stability} adding perturbative noise to a stable pulse for close parameter points.

The TPs have been followed with a pseudo-arclength continuation method \cite{Rheinboldt1988,Mittelmann1986}. The solutions are found using a regular grid with $N=4096$ nodes, $\Delta \xi=0.12 \ su$, and periodic boundary conditions. We have used an integrate-phase condition to choose between the translational equivalent solutions and select the speed $c$ of the TP. 

The stability of the TPs has been determined by a numerical diagonalization of the discretized Jacobian matrix of the TPs in the PDEs in the comoving reference frame. Using the same method we obtained the eigenvalues shown in Fig.~\ref{fig:T_Sneaking}e.

\end{document}